\documentclass[letterpaper]{JHEP3}

\usepackage{amssymb}
\usepackage{amsmath}
\usepackage{graphicx}
\usepackage{bm} 
\usepackage{amssymb}
\usepackage{amstext}
\usepackage{tensor}
\usepackage{epsf,epsfig}
\usepackage{amsfonts} 
\usepackage{graphicx}
 \usepackage{bm} 

\usepackage{amsmath}
\usepackage{mathrsfs}
 
\usepackage{amsmath,amssymb}

\newcommand{\insertplot}[5]{\begin{figure}
 \hfill\hbox to 0.05in{\vbox to #5in{\vfill
 \inputplot{#1}{#4}{#5}}\hfill}
 \hfill\vspace{-.1in}
 \caption{#2}\label{#3}
 \end{figure}}
 \newcommand{\inputplot}[3]{
 \special{ps: plotfile #1}
}
\newcounter{fig}   
\newcommand{\eps}{\epsilon}

\newcommand{\beq}{\begin{equation}}
\newcommand{\eeq}{\end{equation}}
\newcommand{\beqs}{\begin{eqnarray}}
\newcommand{\eeqs}{\end{eqnarray}}

\newcommand{\be}{\begin{equation}}
\newcommand{\ee}{\end{equation}}
\newcommand{\bea}{\begin{eqnarray}}
\newcommand{\eea}{\end{eqnarray}}

\newcommand{\al}{\alpha}

\newcommand{\identity}{{\upright\rlap{1}\kern 2.0pt 1}}

\numberwithin{equation}{section}

\abstract{ 
We study solutions of the Einstein-Skyrme model. Firstly we consider test field Skyrmions on the Kerr background. These configurations -- hereafter dubbed \textit{Skerrmions} -- can be in equilibrium with a Kerr black hole (BH) by virtue of a \textit{synchronisation condition}. We consider two sectors for Skerrmions. In the sector with non-zero baryon charge,
Skerrmions are akin to the known Skyrme solutions on the Schwarzschild background.
These ``topological"
configurations reduce to flat spacetime Skyrmions in a vanishing BH mass limit; moreoever,
they never become ``small" perturbations on the Kerr background:
the non-linearities of the Skyrme model are crucial for all such Skerrmions.
In the non-topological sector, on the other hand, Skerrmions have no analogue on the Schwarzschild background.
Non-topological Skerrmions carry not baryon charge and bifurcate from a subset of Kerr solutions
defining an \textit{existence line}.
Therein the appropriate truncation of the  Skyrme model yield a linear scalar field theory containing a complex plus a real field, both massive and decoupled, and the Skerrmions reduce to the known stationary scalar clouds around Kerr BHs. Moreover, non-topological Skerrmions trivialise in the vanishing BH mass limit.
 We then discuss the backreaction of these Skerrmions, that yield rotating BHs with synchronised Skyrme hair, which continously connect to the Kerr solution (self-gravitating Skyrmions) in the non-topological (topological) sector. In particular, the non-topological hairy BHs provide a non-linear realisation, within the Skyrme model, of the synchronous stationary scalar clouds around Kerr.

}

\preprint{ }

\title{    
Skyrmions around Kerr black holes  and spinning BHs with Skyrme hair  
}

\author{{\large }%$^{\ddagger}$
{\large C. Herdeiro}$^{\dagger,\Diamond}$,
{\large I. Perapechka}$^{\ddagger}$,
{\large E. Radu}$^{\dagger}$,
and {\large Ya. Shnir}$^{\star}$
\\ \\
$^{\dagger}${\small Departamento de F\'isica da Universidade de Aveiro and CIDMA,
}
\\
{\small
 Campus de Santiago, 3810-183 Aveiro, Portugal}
\\
$^{\Diamond}${\small CENTRA,  Departamento  de  F\'\i sica,  Instituto  Superior  T\'ecnico  -  IST,
}
\\
{\small
Universidade  de  Lisboa  -  UL,  Avenida  Rovisco  Pais  1,  1049  Lisboa,  Portugal}
\\
$^{\ddagger}${\small Department of Theoretical Physics and Astrophysics, Belarusian State
University,
Nezavisimosti Avenue 4, Minsk 220004, Belarus}\\% \\
$^{\star}${\small 
BLTP, JINR,
JINR Joliot-Curie 6, Dubna 141980, Moscow Region, Russia}}

\begin{document}

%%%%%%%%%%%%%%%%%%%%%%%%%%%%%%%%%%%%%%%%%%%%%%%%%%%%%%%%%%
\section{Introduction}
%%%%%%%%%%%%%%%%%%%%%%%%%%%%%%%%%%%%%%%%%%%%%%%%%%%%%%%%%%

The legendary simplicity of black holes (BHs), encapsulated in Wheeler's mantra ``BHs have no hair"~\cite{Ruffini:1971bza}
has been challenged by many different models of \textit{hairy BHs} -
see~\cite{Bekenstein:1996pn,Herdeiro:2015waa,Sotiriou:2015pka,Volkov:2016ehx} for recent reviews. Historically, the
first physically relevant counter example to the no-hair conjecture, in the sense of occuring in a sensible physical model
and yielding black holes which are not catastrophically unstable against losing their ``hair", was found in the
$SU(2)$-Skyrme model. This is a sigma model containing four real scalars subject to a constraint. In flat spacetime
its solutions, named \textit{Skyrmions}, provided also the very first explicit example of $solitons$ in a relativistic
non-linear field theory in four spacetime dimensions~\cite{Skyrme:1961vq,Skyrme:1962vh}, and have found interesting applications,
$e.g.$ as an effective description of low energy Quantum Chromodynamics~\cite{Witten:1983tw}
and on the issue of proton decay \cite{Callan:1983nx}. These solutions possess a topological baryon charge being
therefore \textit{topological} solitons.

The Skyrme model started to play a role in the context of ``hairy BHs"  in 1986 when  Luckock and
Moss~\cite{Luckock:1986tr} (see also~\cite{Luckock,Droz:1991cx,Adam:2016vzf,Dvali:2016mur,Gudnason:2016kuu})
observed that a small  BH may be superimposed at the center of a Skyrme soliton, yielding a regular, asymptotically
flat BH solution, non-singular on and outside the horizon - see
also~\cite{Glendenning:1988qy,Droz:1991cx,Bizon:1992gb,Heusler:1991xx,Heusler:1993ci}. These solutions are static and
spherically symmetric.

As found in~\cite{Battye:2005nx}, however, the flat spacetime Skyrmions possess axially symmetric
spinning generalizations\footnote{Spinning Skyrmions were considered already by Skyrme in
his pioneering work \cite{Skyrme:1962vh} in order to identify the quantum numbers of nuclei, assuming the
rigid body approximation.},
which may, moreover, gravitate~\cite{Ioannidou:2006nn} (see also \cite{Perapechka:2017bsb}).
The existence of spinning Skyrmions raises the question if four dimensional asymptotically flat
\textit{spinning} BHs with Skyrme hair exist. One of the purposes of this paper is to answer (positively) this question.

\bigskip

In a parallel development motivated by the linear analysis in \cite{Hod:2012px}, 
it has been shown that asymptotically flat rotating BHs (in various dimensions), can be endowed with ``hair" of different bosonic fields as along as the field obeys a \textit{synchronisation condition}, between the angular velocity of the BH horizon and a phase angular velocity of the matter field. The basic example of this mechanism is provided by Kerr BHs with scalar hair~\cite{Herdeiro:2014goa,Herdeiro:2015gia}. Further examples of fully non-linear solutions were constructed in~\cite{Brihaye:2014nba,Kleihaus:2015iea,Herdeiro:2015kha,Herdeiro:2015tia,Herdeiro:2016tmi,Delgado:2016jxq,Herdeiro:2018wvd}. In some of these cases, the hairy BHs bifurcate from a reference ``bald" solution, such as in~\cite{Herdeiro:2014goa,Kleihaus:2015iea,Herdeiro:2015tia,Herdeiro:2016tmi,Delgado:2016jxq,Herdeiro:2018wvd}, where the reference solution is Kerr BH. In some other cases the hairy BHs are not continuously connected to the reference bald solution, such as~\cite{Brihaye:2014nba,Herdeiro:2015kha}, where the reference BH is the five-dimensional Myers-Perry solution.

The construction of these fully non-linear solutions and, above all, a thorough scanning of their domain of existence is a lengthy and time consuming numerical task. As such, it was suggested in~\cite{Herdeiro:2017oyt} one could probe the existence of these hairy BHs by considering non-linear \textit{$Q$-clouds}~\cite{Herdeiro:2014pka} on the ``bald" BH background. In this way, it could be established, for instance that even black objects with topologically non-spherical horizons, such as black rings~\cite{Emparan:2001wn}, can be endowed with synchronised scalar hair.

$Q$-clouds are a curved spacetime version of the flat spacetime (spinning) \textit{$Q$-ball} solutions~\cite{Coleman:1985ki}, describing a \textit{non-topological} soliton in a model with a single, self-interacting, complex scalar  field~\cite{Volkov:2002aj,Kleihaus:2005me}. Amongst the physical differences between the topological spinning Skyrmions and the non-topological spinning $Q$-balls, we would like to point out that for spinning Skyrmions the angular momentum $J$ is a continuous parameter that can be arbitrarily small, whereas for spinning $Q$-balls it is quantised in terms of their Noether charge, possessing a non-zero minimal value.
Thus the spinning Skyrmions can rotate slowly and rotating configurations are continuously connected to static ones. This will play a role in the discussion below.

\bigskip

Following the strategy of~\cite{Herdeiro:2017oyt} we shall, in this paper, study the existence of Skyrme hair on Kerr BHs, by firstly studying test field Skyrmions ($i.e.$ non-backreacting) on the Kerr background. These solutions, hereafter dubbed \textit{Skerrmions}, exist under the synchronisation condition discussed above. We shall consider Skerrmions that fall into two distinct sectors, according to the existence of a non-vanishing or vanishing baryon number.

In the \textit{topological sector}, Skerrmions reduce to the usual flat spacetime Skyrmions when gravity is switched off.
Such Skerrmions never become ``small", in the sense of becoming solutions of a linear field theory on the Kerr spacetime.
This contrasts to the behaviour of $Q$-clouds, that become linear solutions on an \textit{existence line} of Kerr BHs, wherein the scalar field becomes small, the non-linearities become irrelevant and the $Q$-clouds reduce to the stationary scalar clouds found in~\cite{Hod:2012px} (see also, $e.g.$~\cite{Hod:2014baa,Benone:2014ssa,Wilson-Gerow:2015esa,Bernard:2016wqo,Sakalli:2016xoa,Hod:2016lgi,Ferreira:2017cta}). Such existence line defines the subset of Kerr BHs from which the fully non-linear ``hairy" BH solutions bifurcate. The absence of an existence line anticipates that spinning BHs with topological Skyrme hair do not bifurcate from Kerr BHs, similarly to the aformentioned examples~\cite{Brihaye:2014nba,Herdeiro:2015kha}.
These Skerrmions also possess a non-zero baryon number, which in the limit of vanishing event horizon, becomes an integer -- the topological charge.

In the \textit{non-topological sector},  Skerrmions trivialise in the limit when gravity is switched off.
Such Skerrmions could only be found as linear solutions on an existence line of Kerr BHs, wherein the Skyrme model reduces to a linear field theory composed of a complex together with a real scalar fields, with these fields decoupled and massive,
and carrying no baryon charge. Then, these non-topological Skerrmions, like $Q$-clouds, match the stationary clouds in~\cite{Hod:2012px}, at the existence line. Moreover, the spinning BHs with non-topological Skyrme hair bifurcate from Kerr, similarly to the examples in~\cite{Herdeiro:2014goa,Kleihaus:2015iea,Herdeiro:2015tia,Herdeiro:2016tmi,Delgado:2016jxq,Herdeiro:2018wvd}.

We also construct and discuss some basic properties of backreacting Skerrmions, yielding spinning BHs with Skyrme hair, both in the topological and non-topological sector, in particular verifying the results anticipated by the test field analysis.

This paper is organised as follows. In Section~\ref{section2} we present the basic formalism of the Skyrme model, including the definition of the topological charge, and some comments on the $O(3)$ truncation that yields the non-topological sector. As a preparation for the presentation of Skerrmions, in Section~\ref{section3} we review the spherically symmetric non-backreacting Skyrme solutions on Minkowski as well as those on a Schwarzschild BH, wherein only the topological sector exists. The latter solutions have a branch structure for their global quantities in terms of a dimensionless coupling. In one of the limits of zero coupling the flat spacetime solutions are recovered. In another limit, Yang-Mills solitons are obtained. In Section~\ref{section4} we describe the topological Skerrmion solutions and in~\ref{section5} we discuss the non-topological sector and backreacting Skerrmions that yield spinning BHs with Skyrme hair. Concluding remarks are presented in Section~\ref{section6}.

%%%%%%%%%%%%%%%%%%%%%%%%%%%%%%%%%%%%%%%%%%%%%%%%%%%%%%%%%%
%%%%%%%%%%%%%%%%%%%%%%%%%%%%%%%%%%%%%%%%%%%%%%%%%%%%%%%%%%
\section{The Skyrme model}
\label{section2}
%%%%%%%%%%%%%%%%%%%%%%%%%%%%%%%%%%%%%%%%%%%%%%%%%%%%%%%%%%
%%%%%%%%%%%%%%%%%%%%%%%%%%%%%%%%%%%%%%%%%%%%%%%%%%%%%%%%%%
 Working in four spacetime dimensions with a metric tensor $g_{\mu\nu}$,
we consider the  $SU(2)$-Skyrme  Lagrangian,
 \begin{eqnarray}
\label{LS}
\mathcal{L}_S=
\mbox{Tr}
\left \{
\frac{\kappa^2 }{4}
{\bf K}_{\mu}\,{\bf K}^\mu
 +\frac{1}{32e^2}
   \left[{\bf K}_\mu,{\bf K}_\nu\right]\left[{\bf K}^\mu,{\bf K}^\nu\right]
%
%   F_{\mu \nu}     F^{\mu \nu}
                    \right \}
 +\frac{m_\pi^2}{2}\mbox{Tr}\left \{\frac{{\bf U}+{\bf U}^\dagger}{2}-
\mathbf{1}
%\identity 
\right \},
\end{eqnarray}
where $\kappa$, $e$  and  $m_\pi$ are input parameters\footnote{
In nuclear physics applications, the value of these parameters is fixed
by comparison with experimental data. For the conventions here, the dimensions
of these constants are:
$[\kappa]=1/L$,
$[e]=L^0$
and
$[m_\pi]=1/L^2$.
},
  $m_\pi$ being interpreted as the pion mass.
$ {\bf K}_\mu$ is the $SU(2)$-valued left-invariant current,
\begin{eqnarray}
 {\bf K}_\mu=\partial_\mu {\bf U} {\bf U}^{-1},%~~~{\rm and}~~F_{\mu \nu}=\left[K_\mu,K_\nu\right]
\end{eqnarray}
associated with the
 Skyrme field,
\begin{eqnarray}
\label{Uans}
{\bf U} = \sigma 
\mathbf{1} 
%\identity 
+ i \pi_a {\bf \tau}^a ~~ _{\overrightarrow{r\to \infty}}~~~~
%\identity 
\mathbf{1}
\, ,
\end{eqnarray}
where $\tau^a$ are the Pauli matrices and $\{\sigma, \pi_a \}$ spacetime fields that behave as $\sigma\rightarrow 1$, $\pi_a\rightarrow 0$ in the spatial asymptotic limit $r\rightarrow \infty$.

The Lagrangian~\eqref{LS} contains three terms. The first term is the usual non-linear sigma model term;
 the second is known as `the Skyrme term',
being required by Derrick-type scaling arguments~\cite{Derrick:1964ww} for the existence of finite mass solutions;
the third is a mass term
which is mandatory for rotating solutions, whereas it can be set to zero in the static case.
A generalization of the Skyrme model, which includes an
additional sextic in derivatives term, was suggested recently to construct weakly bounded multisoliton configurations
\cite{Adam:2010fg,Adam:2010ds,Perapechka:2017yyc}.

The Skyrme field satisfies the  equations
\begin{eqnarray}
\label{Ueq}
\nabla_\mu \left(\kappa^2 {\bf K}^\mu
 + \frac{1}{4e^2}[ {\bf K}_\nu, [ {\bf K}^\mu, {\bf K}^\nu]] \right) = \frac{m_\pi^2}{2} ({\bf U}-{\bf U}^\dagger)\ .
\end{eqnarray}
Moreover, the  fields $\{\sigma, \pi_a \}$ which enter the decomposition of the Skyrme field, (\ref{Uans}), are subject to the sigma-model condition,
\begin{eqnarray}
 \sigma^2+ \pi_a  \pi_a =1 \ .
 \label{smc}
\end{eqnarray}
Thus, in the absence of an event horizon,
$\{\sigma, \pi_a \}$ map the compactified coordinate space $S^3$ to the $SU(2)$ group space, which is
isomorphic to the 3-sphere, $S^3$.
The homotopy group of this map is
$\pi_3(S^3)=\mathbb{Z}$; consequently,
each $S^3\longrightarrow S^3$  map falls into an homotopy class indexed by an integer,
which is identified with the \textit{topological charge} $B$:
\begin{equation}
B \equiv  \int_{\Sigma} B^t  d^3x  \ ,
\label{B}
\end{equation}
  %$\Sigma$ is a spacelike surface.
% bounded by the sphere at infinity
% and the horizon,
with the topological current $B^{\mu}$
\begin{equation}
B^\mu \equiv  \frac{1}{\sqrt{-g}} \frac{1}{24\pi^2}
\epsilon^{\mu \nu \alpha \beta} {\rm Tr}\,
\left( {\bf K}_\nu {\bf K}_\alpha {\bf K}_\beta \right) \ .
\end{equation}
The (static) solution found by Skyrme using numerical methods~\cite{Skyrme:1961vq,Skyrme:1962vh} and whose existence proof was given in \cite{Esteban:1986dm}, has $B=1$. We emphasize that the topological charge
$B$ does not arise from an invariance of the matter
Lagrangian under any symmetry transformation and is therefore $not$ a Noether
charge \cite{Manton:2004tk}, unlike the charge of $Q$-balls. Its origin comes  instead from the non-trivial topology of the Skyrmions.

In the case of the BHs with Skyrme hair, strictly speaking, the topology ($i.e.$ the mapping between physical and internal spaces)
is lost. Still, one defines a  baryon charge $B$ by performing the integration~\eqref{B} in the exterior region. The numerical results in~\cite{Glendenning:1988qy,Droz:1991cx,Bizon:1992gb,Heusler:1991xx,Heusler:1993ci} show that the  BH  horizon ``absorbs" a part of the baryon charge, but $B$ never vanishes for a static horizon and thus the vacuum BH is never approached. As we shall see below, the same holds for the topological sector of Skerrmions.

In the non-topological sector, on the other hand, corresponding to a particular $O(3)$ truncation of the full Skyrme model, Skerrmions always have $B=0$. In this case, moreover, Skerrmions become infinitesimally small. To better understand this point, let us consider small perturbations around the Skyrme vacuum state (thus $B=0$, trivially\footnote{
The Skyrme model possesses
solutions with
$B=0$ and nonzero mass,
describing Skyrme-anti-Skyrme composite states \cite{Krusch:2004uf,Shnir:2009ct}
(likely unstable). These solutions can be studied within the same Ansatz employed in this work. They satisfy, however, a different set of boundary conditions and they shall not be considered in this work.
}); these perturbations are described by  parameterising  the $\{\sigma, \pi_a \}$ fields as
\begin{eqnarray}
\label{s1}
\pi_a=\eps \Pi_a \ ,~~ \sigma=1+\eps^2 \Sigma\ ,
\end{eqnarray}
where $\epsilon\ll 1$ and $\Sigma=- \Pi_a \Pi_a/2$ from the sigma-model condition~\eqref{smc}.
Then, defining the complex scalar field $\psi$ and a real one  $\chi$ as
\begin{eqnarray}
\label{s3}
 \psi=\Pi_1 +i \Pi_2,~~~\chi=\Pi_3 \ ,
\end{eqnarray}
to leading order, the Skyrme Lagrangian~\eqref{LS} reduces to  $\mathcal{L}_{S}\simeq \mathcal{L}_\psi+\mathcal{L}_{\chi}$, where
\begin{eqnarray}
\label{s2}
{\mathcal L}_\psi=-\partial_\mu \psi^{*}\partial^\mu \psi-m_\pi^2 \psi^{*}  \psi \ , \qquad
{\mathcal L}_{\chi}=-\partial_\mu \chi\partial^\mu \chi-m_\pi^2 \chi^2 \ .
\end{eqnarray}

On a static BH,
both $\psi$ and $\chi$
vanish identically.
However, the presence of a rotating horizon allows for nontrivial fields $\psi,\chi$
 forming bound states for a particular set of Kerr backgrounds. These correspond precisely to the existence line discussed in~\cite{Hod:2012px,Herdeiro:2014goa}. Thus, at the linear  level, stationary scalar clouds of a complex massive scalar field are solutions of the Skyrme model  on a Kerr backround. At the non-linear fully backreacting model, the Skyrme model, in the non-topological sector, will provides a more fundamental context wherein hairy BHs supported by the synchronisation condition arise, akin to Kerr BHs with scalar hair~\cite{Herdeiro:2014goa}. We remark that for these nonlinear realizations of the complex field $\psi$,
the real scalar field
$\chi$
(which is infinitesimally small on the Kerr existence line)
must vanishes identically to all orders.
Otherwise, the energy-momentum tensor
of the real scalar field $\chi$ would  possess a dependence on both time and azimuthal coordinates and its backreacting version would not be compatible with a stationary, axi-symmetric geometry.
%

%%%%%%%%%%%%%%%%%%%%%%%%%%%%%%%%%%%%%%%%%%%%%%%%%%%%%%%%%%
%%%%%%%%%%%%%%%%%%%%%%%%%%%%%%%%%%%%%%%%%%%%%%%%%%%%%%%%%%
\section{Spherical solutions}
\label{section3}
%%%%%%%%%%%%%%%%%%%%%%%%%%%%%%%%%%%%%%%%%%%%%%%%%%%%%%%%%%
%%%%%%%%%%%%%%%%%%%%%%%%%%%%%%%%%%%%%%%%%%%%%%%%%%%%%%%%%%

As a preparation for Skerrmions let us start by reviewing spherically symmetric solutions. Besides being much simpler,  they
possess already a number of basic properties of the corresponding rotating generalizations in the topological sector.

Working in spherical coordinates, we consider a general metric Ansatz:
\begin{eqnarray}
\label{sph}
ds^2=-f_0(r)dt^2+f_1(r)[dr^2+r^2(d\theta^2+\sin^2 d\varphi^2)] \ .
\end{eqnarray}
For the Skyrme field, we take the usual
  spherically symmetric Skyrme ansatz~\eqref{Uans}:\footnote{Strictly speaking the Skyrme field
    (\ref{sph1})
    is not
      spherically symmetric.
That is, the effect of a spatial rotation of the Skyrme field can
be compensated by an isospin transformation, rather than the Skyrme field being
only dependent on the radial coordinate.
However, both the
  energy density and baryon charge density are spherically symmetric,
    the Ansatz (\ref{sph1})
    being compatible with the line element (\ref{sph}). Thus, one may regard these solutions as another case of symmetry non-inheritance~\cite{Smolic:2015txa}.}
\begin{eqnarray}
\label{sph1}
\sigma= \cos F(r)\ , \qquad \pi_a=n_a \sin F(r)  \ ,
\end{eqnarray}
where $n_a$ are the components of the unit vector:
\begin{eqnarray}
\label{sph2}
n_a=(\sin \theta \cos m\varphi,\sin \theta \sin m\varphi,\cos \theta  )
\end{eqnarray}
 and $m$ is a positive integer, $m\in \mathbb{N}$.
Thus, in the probe limit, the problem
reduces to solving a single ordinary differential equation
for the Skyrme function $F(r)$. In what follows we shall restrict our study to $m=1$
solutions.

For both Minkowski and Schwarzschild Skyrmions,  the total mass of the solutions in the probe limit, $M$, is computed
as the integral
of the $T_t^t$ component  of the energy momentum tensor (as given by (\ref{Tik}) below), while the topological charge,~\eqref{B},  for
the Skyrme Ansatz (\ref{sph1}), reduces to
\begin{eqnarray}
\label{32w}
 B=\frac{1}{\pi}
\left\{
-F(r)+\frac{1}{2}\sin [2F(r)]
\right\}
\bigg|_{r_0}^\infty \ .
\end{eqnarray}
For Minkowski Skyrmions, $r_0=0$; for solutions on Schwarzschild  BH background $r_0=r_H$.

%%%%%%%%%%%%%%%%%%%%%%%%%%%%%%%%%%%%%%%%%%%%%%%%%%%%%%%%%%
\subsection{Minkowski Skyrmions}
%%%%%%%%%%%%%%%%%%%%%%%%%%%%%%%%%%%%%%%%%%%%%%%%%%%%%%%%%%
For a Minkowski background, $f_0(r)=f_1(r)=1$ in~\eqref{sph}. Then, the near origin solution for $F(r)$ is:
\begin{eqnarray}
\label{sph4}
F(r)=\pi-b r^2+\mathcal{O}(r^4) \ ,
\end{eqnarray}
while the far field expression reads
\begin{eqnarray}
\label{sph5}
F(r)=\frac{e^{-m_\pi r}}{r}+\dots \ .
\end{eqnarray}
Thus, from (\ref{32w}), the topological charge of the horizonless spherical
Skyrmion with $m=1$ is $B=1$.

The profile of the (fundamental, flat space) Skyrme soliton is shown in Figure 1.
One can see that $F(r)$ interpolates smoothly between $\pi$
and $0$, without nodes.
Here we use the same scaling and units as in the case of a
Schwarzschild BH background (as discussed below).

%%%%%%%%%%%%%%%%%%%%%%%%%%%%%%%%%%%%%%%%%%%%%%%%%%%%%%%%%%%%%%%%
\begin{figure}[ht!]
%\lbfig{rhfar}
\begin{center}
\includegraphics[height=.34\textwidth, angle =0 ]{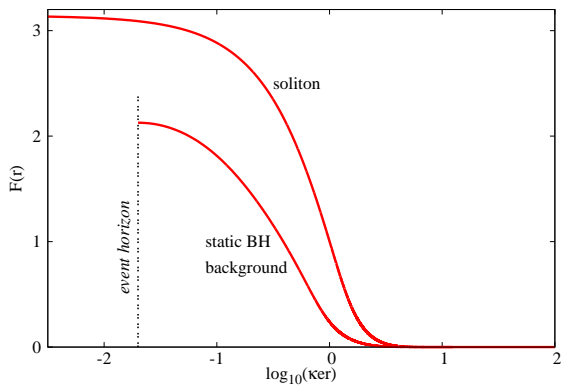}
\end{center}
\caption{
 The profile of the Skyrme function,  $F(r)$,
% is shown as a function of the dimensionless radial coordinate $\kappa e r$, 
for the
fundamental Minkowski and Schwarzschild Skyrmions. 
The mass term vanishes in  both cases, $m_\pi=0$, while $r_H \kappa e = 0.02$.
}
\label{profile-spherical}
\end{figure}
%%%%%%%%%%%%%%%%%%%%%%%%%%%%%%%%%%%%%%%%%%%

%%%%%%%%%%%%%%%%%%%%%%%%%%%%%%%%%%%%%%%%%%%%%%%%%%%%%%%%%%
\subsection{Schwarzschild Skyrmions}
%%%%%%%%%%%%%%%%%%%%%%%%%%%%%%%%%%%%%%%%%%%%%%%%%%%%%%%%%%

Considering isotropic coordinates,
the Schwarzschild BH can be written in the form (\ref{sph})
with
\begin{eqnarray}
 f_0(r)=
\frac{(1-\frac{r_H}{r})^2}
{(1+\frac{r_H}{r})^2} \ , \qquad   f_1(r)=\left(1+\frac{r_H}{r}\right)^4
 \ ,
\end{eqnarray}
with the BH mass, $M_{Sch}$,  being related to the event horizon radius in this coordinates, $r_H$, as
\begin{eqnarray}
M_{Sch}=2 r_H \ .
\end{eqnarray}
In this case there is a new length scale, $r_H$.

The Skyrme field Ansatz
is still given by (\ref{sph1}). Then we obtain the effective Lagrangian:
\begin{eqnarray}
\mathcal{L}_{\rm eff}=r^2\sqrt{f_0f_1^3}
\left[
\frac{\kappa^2}{2f_1}\left(F'^2+\frac{2\sin^2F}{r^2}\right)
+\frac{1}{e^2} \frac{2\sin^2F}{r^2 f_1^2}
\left(
F'^2+\frac{2\sin^2F}{2r^2}
\right)
+m_\pi^2(1-\cos F)
\right] \ .
\end{eqnarray}
While the far field expression is still given by
(\ref{sph5}),
the near horizon form of the solution is
\begin{equation}
\label{nh1}
F(r)=F_H+F_2(r-r_H)^2+\dots, ~~ {\rm with} ~~
F_2=\frac{\sin F_H \cos F_H}{4 r_H^2}+
\frac{(16 m_{\pi}^2 r_H^2+\kappa^2 \cos F_H)\sin F_H}{4\kappa^2  r_H^2+\sin^2 F_H/e^2 }\ .
\end{equation}
The corresponding value of the baryon charge is
\begin{eqnarray}
B=\frac{1}{\pi}
\left(
F_H-\sin F_H \cos F_H
\right).
\end{eqnarray}

In Figure~\ref{profile-spherical} we also show the profile of a typical Skyrmion
on a Schwarzschild BH background. The Skyrme function $F(r)$
is still monotonic, but $F(r_H)\neq \pi$. Observe that $B\neq  0$ and thus these Schwarzschild Skyrmions belong to the topological sector (there is no non-trivial non-topological sector on Schwarzschild).

 %%%%%%%%%%%%%%%%%%%%%%%%%%%%%%%%%%%%%%%%%%%%%%%%%%%%%%%%%%
\subsubsection{Branch structure and the Yang-Mills limit}
%%%%%%%%%%%%%%%%%%%%%%%%%%%%%%%%%%%%%%%%%%%%%%%%%%%%%%%%%%
To understand how the mass and baryon charge of Skyrmions on Schwarzschild vary in the space of solutions, let us remark on scaling and units. The Skyrme model has a natural length scale in terms of the input constants of the model,  $L_S\equiv {1}/{\kappa e}$.
Then, we define a dimensionless radial coordinate, $x$, as $x\equiv {r}/{L_S}=\kappa e r$.
For the BH backgrounds, we define the dimensionless ``coupling constant"
\begin{eqnarray}
\alpha^2\equiv \frac{M_{\rm Schw}}{L_S}=2r_H \kappa e \ .
\end{eqnarray}
One defines also a dimensionless pion mass
%\bar m_\pi=\frac{m_\pi}{\kappa^2 e}
$\mu\equiv {m_\pi}/({\kappa^2 e})$,
while the
relation between the (displayed) numerical value for the total mass-energy for Skyrmions, $M^{(num)}$,
 and the
 dimensionful one $M$,  is
\begin{eqnarray}
M^{(num)}=M\frac{\kappa}{e} \ .
\end{eqnarray}

%%%%%%%%%%%%%%%%%%%%%%%%%%%%%%%%%%%%%%%%%%%%%%%%%%%%%%%%%%%%%%%%
\begin{figure}[ht!]
%\lbfig{rhfar}
\begin{center}
\includegraphics[height=.34\textwidth, angle =0 ]{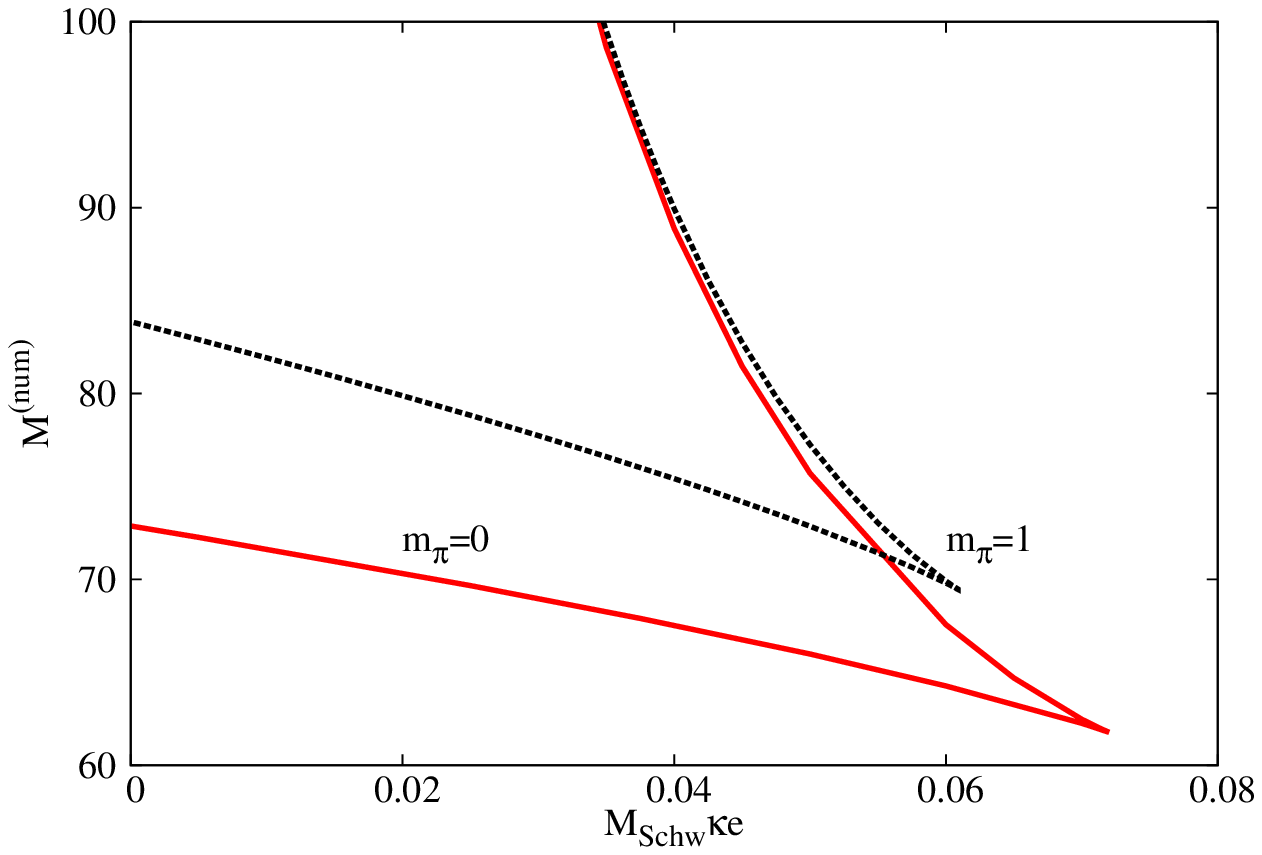}
\includegraphics[height=.34\textwidth, angle =0 ]{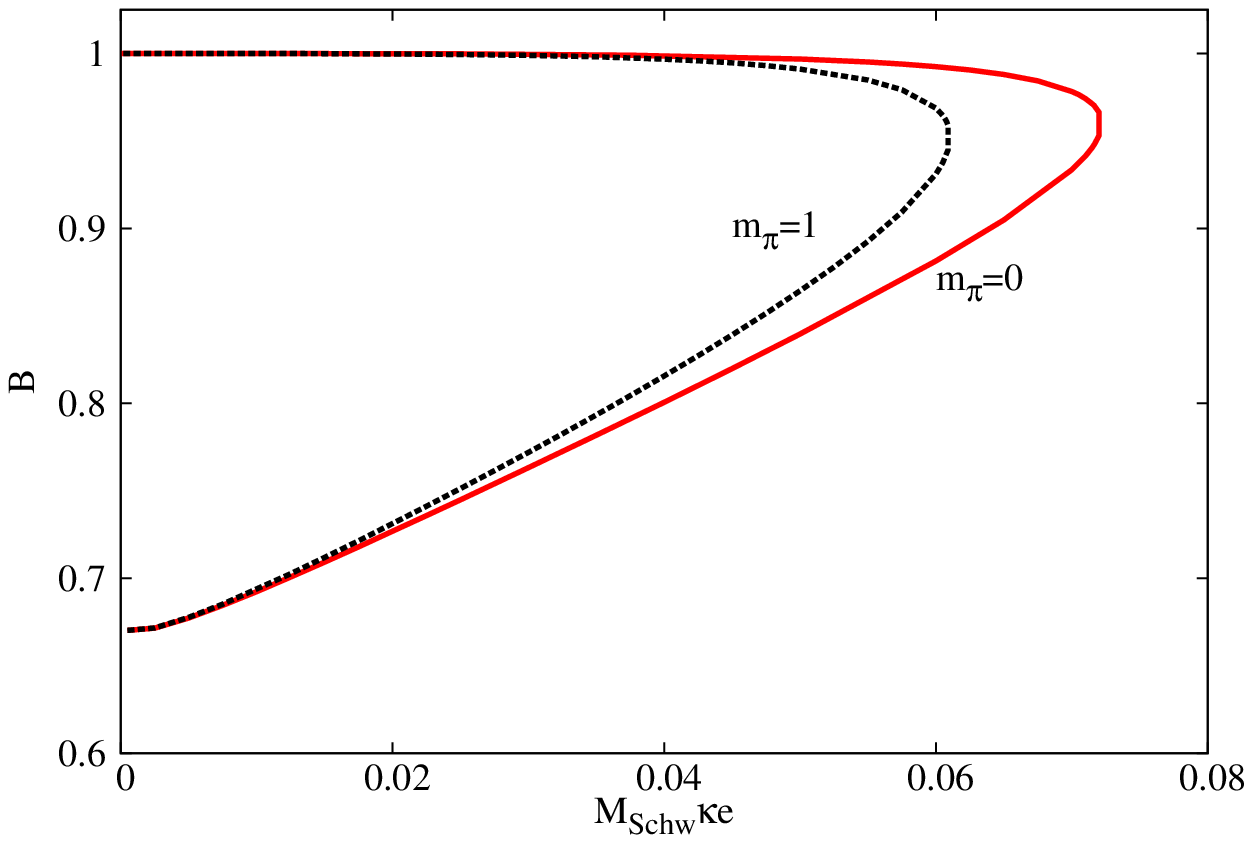}
\end{center}
\caption{
 Mass (left panel) and baryon charge (right panel) for
Skyrmions on a Schwarzschild BH, in terms of the coupling $\alpha=M_{\rm Schw} \kappa e$, for two illustrative values of the pion mass.
}
\label{Schw2}
\end{figure}
%%%%%%%%%%%%%%%%%%%%%%%%%%%%%%%%%%%%%%%%%%%

The variation of the Skyrmions mass $M$ and baryon charge $B$
 with the coupling constant $\alpha$
is shown  in Figure \ref{Schw2} for illustrative values of the pion mass.
Given a set of input parameters
in the Skyrme Lagrangian, BHs able to support Skyrmions cannot be arbitrarily large,
with the existence of a maximal value for the BH mass.
Moreover, for any allowed value of $\alpha$,
one notices the existence of two branches of solutions
which merge for the maximal $\alpha$.
The baryon charge $B$ decreases along both branches, approaching the minimal value
for the limiting solution on the second branch.

Due to the branch structure, there are two ways to approach the limit
$\alpha \to 0$.
Firstly, it
 can be approached as $r_H\to 0$  ($i.e.$ a flat spacetime background).
%or as $\kappa \to 0$,
%($i.e.$ no quadratic term in the Skyrme action).
These are the solutions which form  the fundamental branch (since they
can be smoothly connected to the flat space Skyrmions).
They are of special interest, being stable.
Secondly, it can be approached along a branch  of excited 
configurations which emerge from the fundamental branch, at the maximal value of $\alpha$.
In this case, the limit $\alpha \to 0$ is approached as  $\kappa \to 0$, corresponding to the absence of the
quadratic term in the Skyrme Lagrangian. 
The  Skyrmion mass $M^{(num)}$   diverges in this limit, but the product $\alpha M^{(num)}$
approaches a finite value independent of $\mu$.\footnote{An interesting different
branch structure is observed in the modified twelfth-order Einstein-Skyrme model \cite{Gudnason:2018lhn}.}

In order
to understand this behaviour we introduce a new dimensionless radial coordinate $y$, with
 \begin{eqnarray}
\label{newrad}
\bar r= r_H \kappa y~~~i.e.~~r=r_H e y \ ,
\end{eqnarray}
the length scale being fixed here by the horizon radius $r_H$.
Then, after taking the limit $\kappa \to 0$
  (with fixed  $\mu$)
one finds that the system is described by the  reduced effective Lagrangian
\begin{eqnarray}
\label{YM}
\mathcal{L}_{\rm eff}= 4\sqrt{\frac{f_0}{f_1 }}
\left[
\left(\frac{dw_0}{dy}\right)^2+\frac{(1-w_0^2)^2 }{2y^2}
\right] \ ,
\end{eqnarray}
with $w_0(y)=\cos F(y)$.
One recognizes
(\ref{YM})
as  corresponding to the
  $SU(2)$-Yang-Mills (YM) effective Lagrangian.
%(with a field strength tensor $F_{\mu \nu}=\left[K_\mu,K_\nu\right]$,
Here one considers the usual YM spherical connection with $w_0$ corresponding to the magnetic gauge potential
(and a vanishing electric potential).
Then the limiting solution corresponds to
a spherically symmetric YM
configuration on a fixed Schwarzschild background.
The (fundamental) exact solution has been found long time ago~\cite{BoutalebJoutei:1979va},\footnote{Historically, this largely unknown paper, describes the first example of hair on Schwarzschild BH (in the probe limit).}
its expression in isotropic coordinates being
%(to simplify the notation, here we return to usual radial coordinate)
\begin{eqnarray}
\label{exact-sol}
w_0(r)=-\frac{\sqrt{3}+\frac{1-\sqrt{3}}{4 r r_H}(r+r_H)^2}
{3-\frac{1-\sqrt{3}}{4 r r_H}(r+r_H)^2} \ .
\end{eqnarray}
From (\ref{32w}), this implies a value of the baryon charge
$B \simeq 0.668518$
which provides a minimum for all solutions.
The total mass in this limit (after multiplying with $\kappa$) is
\begin{eqnarray}
\label{Mlim}
M \simeq 16\pi M_{Schw} 0.9587.
\end{eqnarray}
These results are independent of the presence of a mass term in the Skyrme Lagrangian.

%%%%%%%%%%%%%%%%%%%%%%%%%%%%%%%%%%%%%%%%%%%%%%%%%%%%%%%%%%
%%%%%%%%%%%%%%%%%%%%%%%%%%%%%%%%%%%%%%%%%%%%%%%%%%%%%%%%%%
\section{Topological Skerrmions}
\label{section4}
%%%%%%%%%%%%%%%%%%%%%%%%%%%%%%%%%%%%%%%%%%%%%%%%%%%%%%%%%%
%%%%%%%%%%%%%%%%%%%%%%%%%%%%%%%%%%%%%%%%%%%%%%%%%%%%%%%%%%

We now turn our attention to the solutions of the Skyrme model on a fixed Kerr BH geometry with non-trivial baryon charge (topological sector). In quasi-isotropic coordinates, the Kerr line element is
\begin{equation}
\label{metric}
ds^2 = -F_0(r,\theta) dt^2
       +F_1(r,\theta) \left(dr^2+r^2 d\theta^2\right)
       +F_2(r,\theta) r^2 \sin^2\theta
          \left[d\varphi-W(r,\theta) dt\right]^2 \  ,
\end{equation}
where 
\cite{Brandt:1996si}
\begin{eqnarray}
\label{Kerr}
&&
F_1(r,\theta)=\frac{2M_{\rm Kerr}^2}{r^2}+\left(1-\frac{r_H^2}{r^2}\right)^2+
\frac{2M_{\rm Kerr}}{r}\left(1+\frac{r_H^2}{r^2}\right)-\frac{M_{\rm Kerr}^2-4r_H^2}{r^2}\sin^2\theta\ ,~~
\\
&&
\nonumber
F_2(r,\theta)=\frac{S(r,\theta)}{F_1(r,\theta)} ,
~~
F_0(r,\theta)=\left(1-\frac{r_H^2}{r^2}\right)^2\frac{F_1(r,\theta)}{S(r,\theta)},~~
\\
&&
\nonumber
W(r,\theta)=\frac{2M_{Kerr}\sqrt{M_{\rm Kerr}^2-4r_H^2}}{r^3}\frac{1+\frac{M_{\rm Kerr}}{r}+\frac{r_H^2}{r^2}}{S(r,\theta)},
\end{eqnarray}
with
\begin{eqnarray}
\nonumber
S(r,\theta)=
\left[\frac{2M_{\rm Kerr}^2}{r^2}+\left(1-\frac{r_H^2}{r^2}\right)^2+\frac{2M_{\rm Kerr}}{r}\left(1+\frac{r_H^2}{r^2}\right)\right]^2
-\left(1-\frac{r_H^2}{r^2}\right)^2\frac{M_{\rm Kerr}^2-4r_H^2}{r^2}\sin^2\theta~,
\end{eqnarray}
and contains two input parameters, the event horizon radius $r_H$
and the BH mass $M_{\rm Kerr}$. The event horizon angular velocity is determined by these parameters:
\begin{eqnarray}
\label{OmegaH}
\Omega_H=\frac{\sqrt{M_{\rm Kerr}^2 -4r_H^2}}{2M_{\rm Kerr}(M_{\rm Kerr}+2r_H) }  \ .
\end{eqnarray}

The fields $\{\sigma, \pi_a \}$ in the Skyrme field Ansatz~\eqref{Uans}, are defined by three real functions $\phi_a$, subject to a constaint~\cite{Battye:2005nx,Ioannidou:2006nn},
\begin{eqnarray}
\label{U}
 \pi_1+i \pi_2=\phi_1(r,\theta)e^{i(m\varphi-w t)}\ , \ \ \ \ 
 \pi_3=\phi_2(r,\theta) \ , \ \ \ \  \sigma=\phi_3(r,\theta) \ , ~{\rm with}~ \sum_{a=1}^3 \phi_a \phi_a=1\ ,
\end{eqnarray}
where $w$ is the harmonic frequency of the Skyrme field. The spherical limit is recovered for $\phi_a=n_a F(r)$.

The solutions are found again by solving a boundary value problem.
The boundary conditions imposed at spatial infinity are $\phi_1 \big|_{r=\infty}=\phi_2 \big|_{r=\infty} =0,~~\phi_3 \big|_{r=\infty} =1,$
while on the symmetry axis we impose\footnote{
These boundary conditions result from an approximate construction of the solution  on the symmetry axis  
as a power series in $\theta$ (or $\pi-\theta$)
and are compatible with the finiteness of $T_t^t$ and $T_\varphi^t$-components of the energy-momentum tensor.
}
$
\phi_1 \big|_{\theta=0,\pi}=
\partial_\theta \phi_2 \big|_{\theta=0,\pi}=
\partial_\theta \phi_3 \big|_{\theta=0,\pi}=0.
$
Of special interest is the behaviour as $r\to r_H$.
Assuming the existence of a power series expansion in $r-r_H$,
 first one finds the following {\it synchronisation condition}
between the field frequency and the event horizon angular velocity~\eqref{OmegaH}.
\begin{eqnarray}
\label{synch}
w= m \Omega_H~ .
\end{eqnarray}
This condition implies that there is no net flux of the Skyrme field
across the horizon and also that the components
$T_t^t$ and $T_\varphi^t$
of the energy-momentum tensor
are finite at the horizon.
Second, the same approximate solution in $r-r_H$ 
implies the following horizon boundary conditions:
$
\partial_r \phi_1 \big|_{r=r_H}=\partial_r \phi_2 \big|_{r=r_H}=\partial_r \phi_3 \big|_{r=r_H}=0,
$
which are imposed in the numerical procedure.
Also, all solutions in this work possess a reflection symmetry along the equatorial plane
($\theta=\pi/2$),
which implies\footnote{The imposed conditions 
on the boundaries of the domain of integration
are compatible with
those imposed
 both in the static limit
(see (\ref{sph}) 
and
(\ref{sph5}), (\ref{nh1})) and in  
the spinning, solitonic limit of the solutions (as given $e.g.$ in  \cite{Ioannidou:2006nn}).
} 
$
\partial_\theta \phi_1 \big|_{\theta=\pi/2}=
 \phi_2 \big|_{\theta=\pi/2}=
\partial_\theta \phi_3 \big|_{\theta=\pi/2}=0. 
$

For the Ansatz (\ref{U}), the baryon charge is 
\begin{eqnarray}
\label{Qtop}
B= \frac{m}{\pi} \int_{r_H}^\infty dr \int_0^\pi d\theta
\rho(r,\theta),~~{\rm with}~~
\rho(r,\theta)=\partial_r \phi_2 \partial_\theta \phi_3
-
 \partial_r \phi_3 \partial_\theta \phi_2 \  .
\end{eqnarray}

Skerrmion solutions are
found by using the same scaling and units as in the static limit,
together with
\begin{eqnarray}
 w\to w \kappa e\ , \qquad \Omega_H \to \Omega_H \kappa e \ .
\end{eqnarray}

Most of the non-spherical solutions\footnote{Some
of the solutions were also constructed by using a 
 professional package \cite{schoen},
which also uses the Newton-Raphson method.
We have found
a very good agreement between the results obtained within these different approaches.
}
reported
in this work were found by
using a self-implemented finite difference method
based on Newton-Raphson method with PARDISO as a linear solver
\cite{pardiso}.
In numerics,
a compactified radial coordinate $x$ was introduced  with
$x=(r-r_H)/(1+r)$.
Most of the results have been obtained for
 an equidistant grid in $(x,\theta)$, with $100 \times 50$ points.
The sigma-model constraint $\sum_{a=1}^3 \phi_a \phi_a=1$
is imposed by using the Lagrange multiplier method,
as explained $e.g.$ in \cite{Rajaraman,Radu:2008pp}.

%%%%%%%%%%%%%%%%%%%%%%%%%%%%%%%%%%%%%%%%%%%%%%%%%%%%%%%%%%%%%%%%
\begin{figure}[ht!]
%\lbfig{rhfar}
\begin{center}
\includegraphics[height=.34\textwidth, angle =0 ]{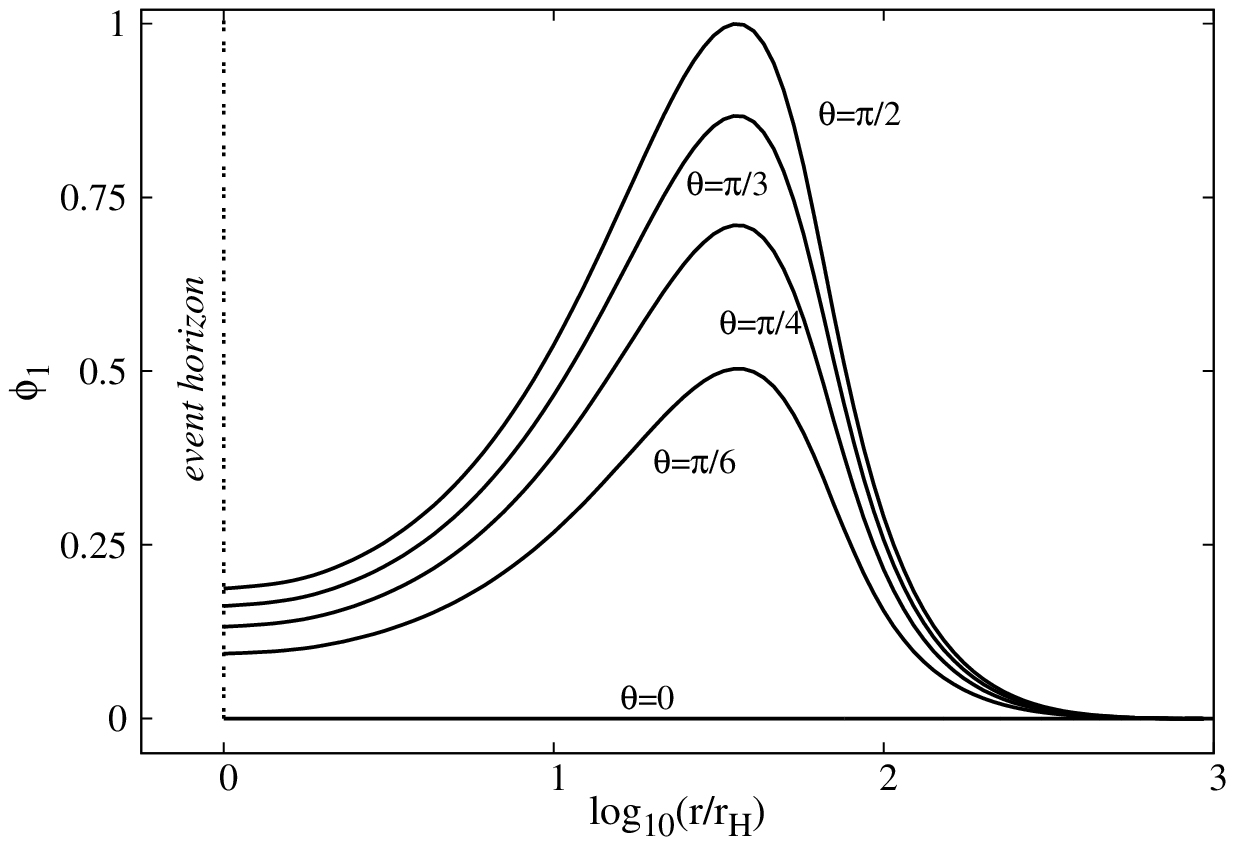}
\includegraphics[height=.34\textwidth, angle =0 ]{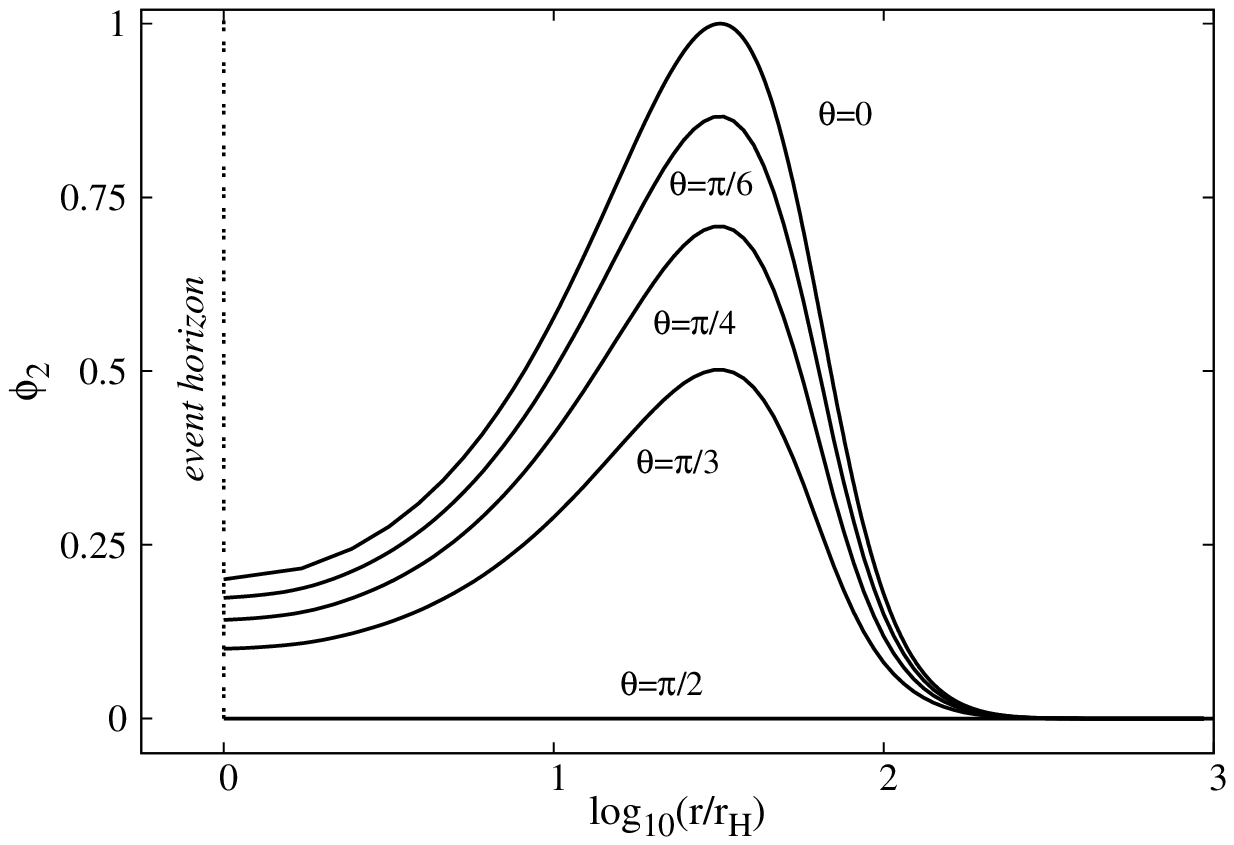}
\includegraphics[height=.34\textwidth, angle =0 ]{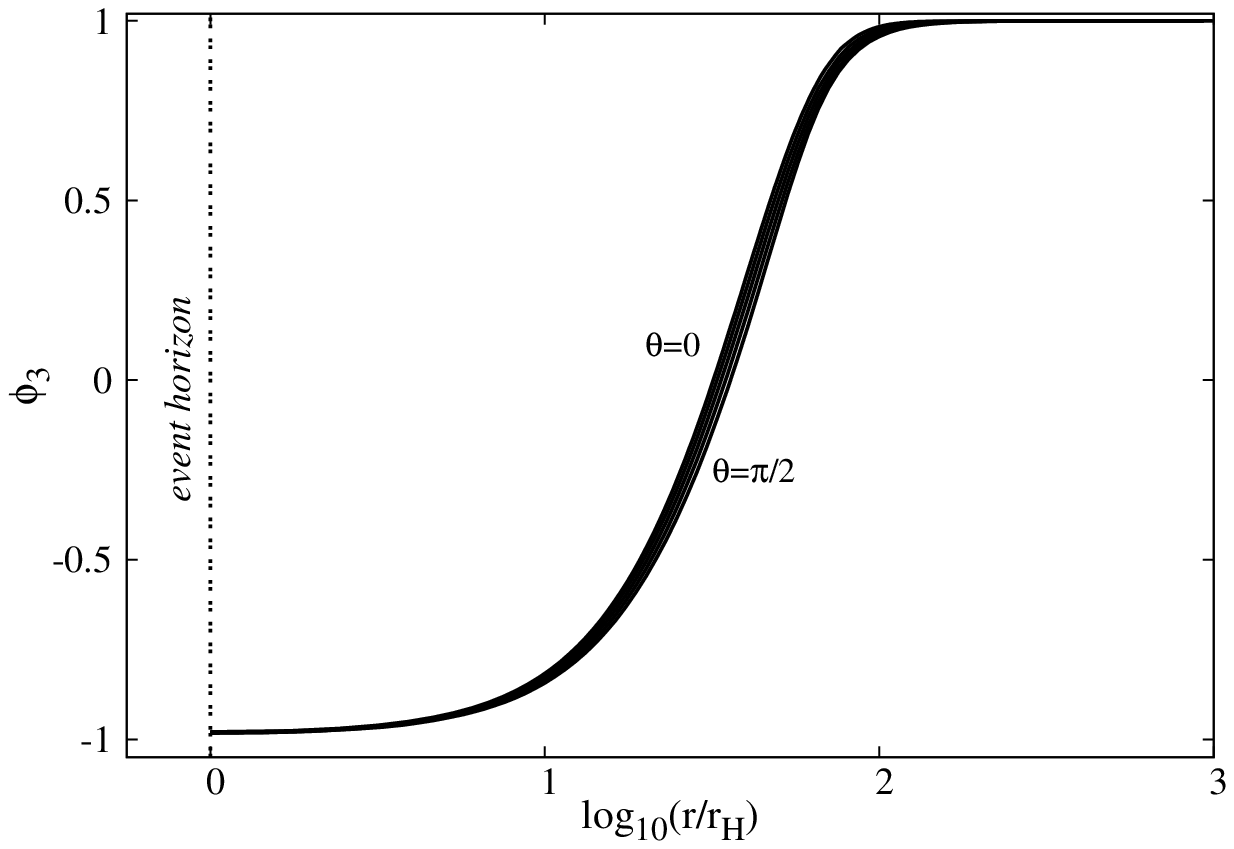}
\includegraphics[height=.34\textwidth, angle =0 ]{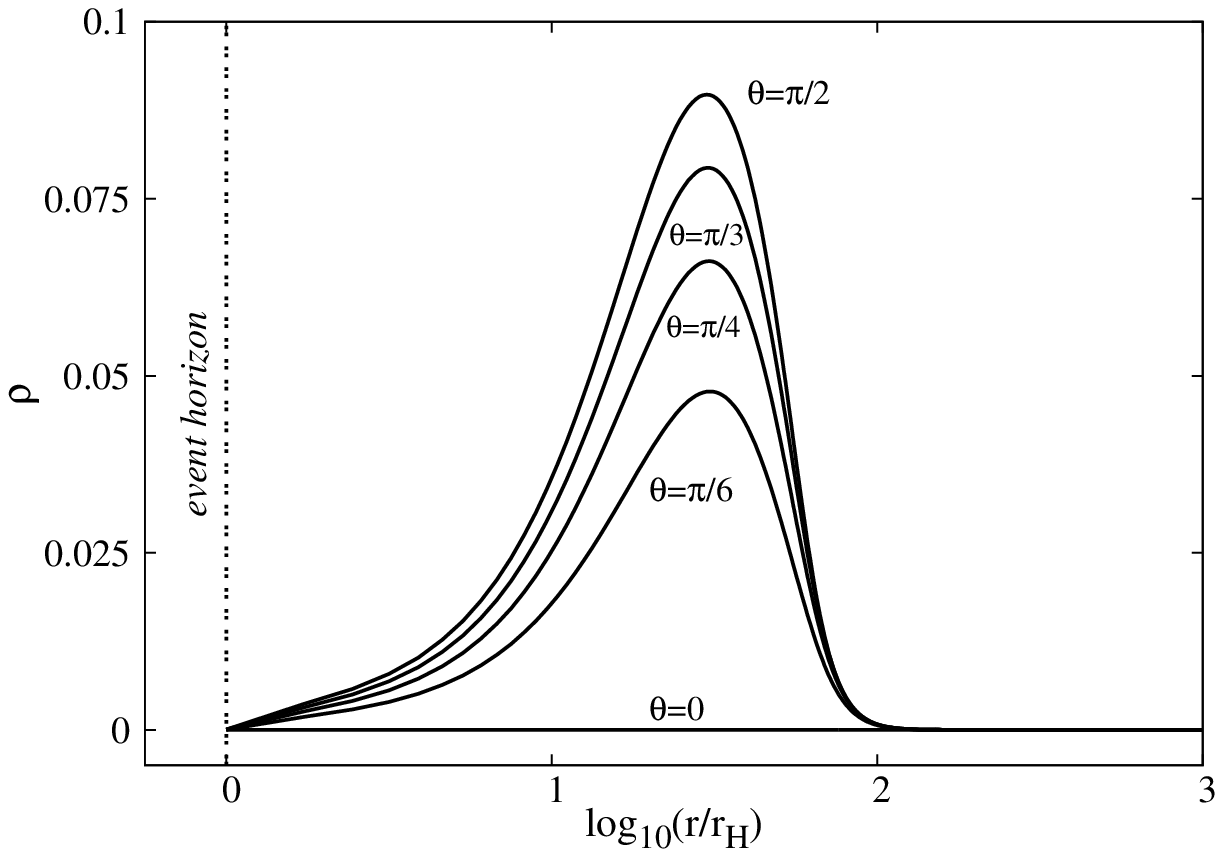}
\includegraphics[height=.34\textwidth, angle =0 ]{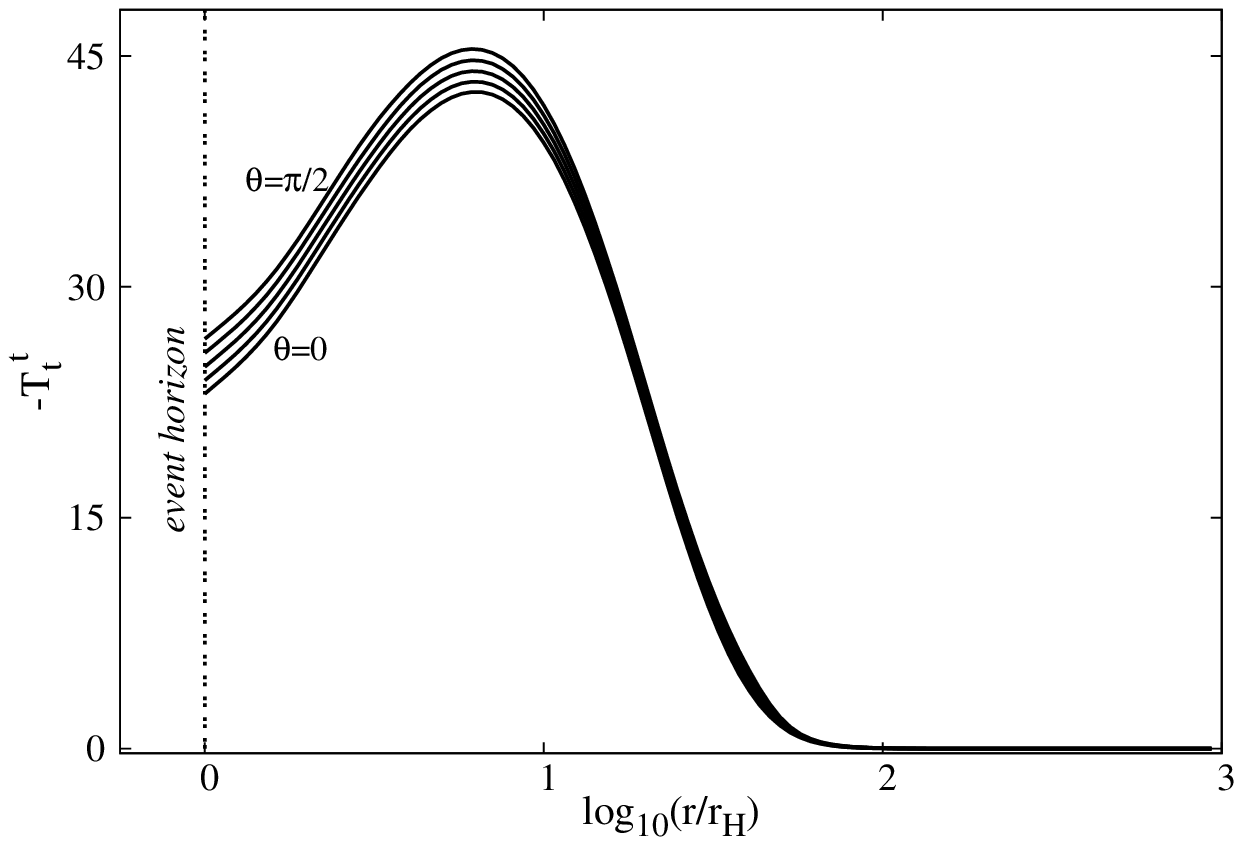}
\includegraphics[height=.34\textwidth, angle =0 ]{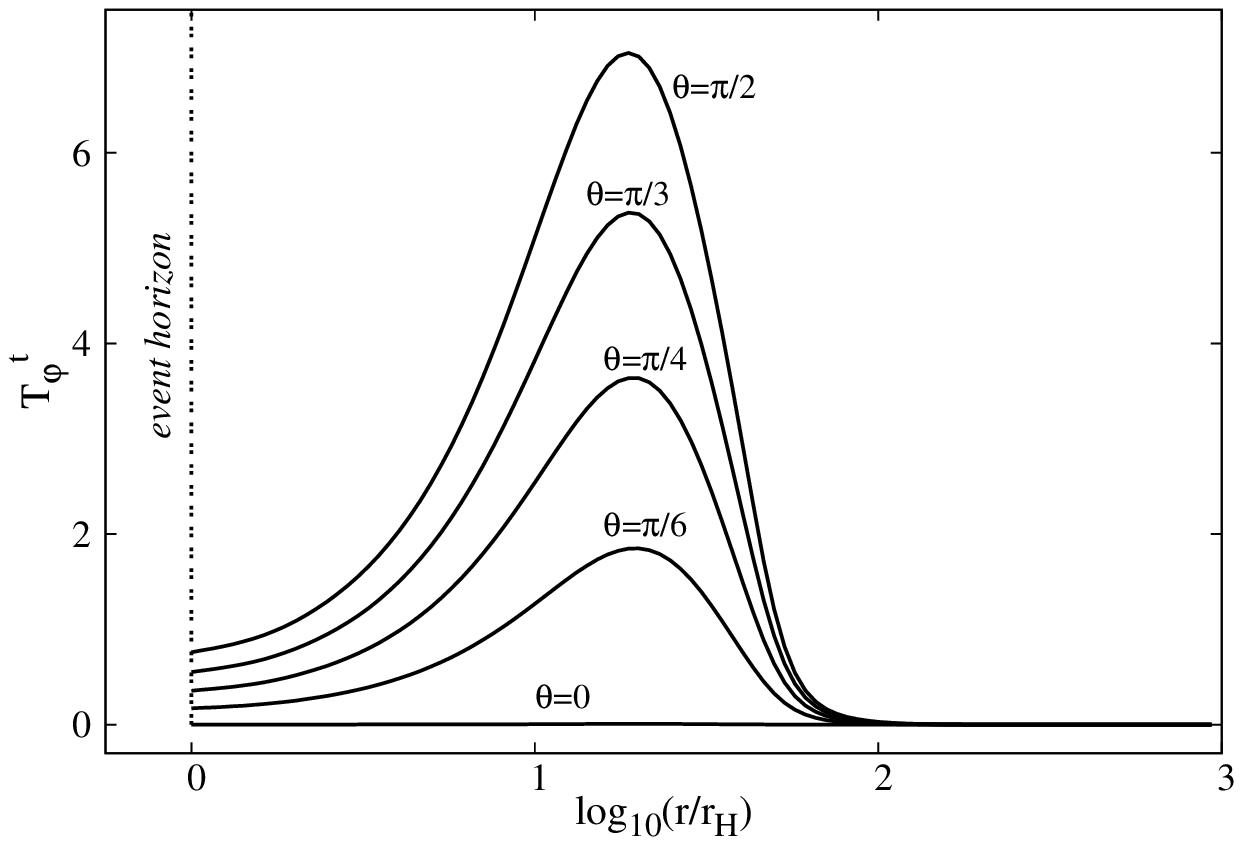}
\end{center}
\caption{
 The  functions $\phi_a$
and the baryon charge density $\rho$, energy and angular momentum densities
are shown for Skerrmions on a Kerr background with $M_{\rm Kerr}=0.04$ and $\Omega_H=0.95$.
}
\label{profileKerr}
\end{figure}
%%%%%%%%%%%%%%%%%%%%%%%%%%%%%%%%%%%%%%%%%%%

 The profile of a typical solution is shown in Figure \ref{profileKerr}, using a choice of the (scaled) pion mass $\mu=1$ and winding number $m=1$. One can see that
$e.g.$
the deformation from sphericity of the
energy density
is rather small, while the angular momentum density is localized in a torus around the equatorial plane.
The properties of these topological Skerrmions are summarised in the following.

Firstly, we have observed that any (spinning) solitonic solution
appears to possess generalizations with a small horizon at its center,
provided the Skyrme field and the horizon are in synchronous rotation.
The field frequency ($i.e.$ the event horizon angular velocity)
cannot be arbitrary large.
A linearization in the far-field  of the field equation for $\psi=\pi_1+i \pi_2$
yields
\begin{eqnarray}
%\left(
% \frac{1}{r^2 \sin }\frac{\partial }{\partial r}(r^2 \frac{\partial }{\partial r})
(\nabla_0^2
+w^2-\mu^2
 )\psi=0 \ ,
\end{eqnarray}
where $\nabla_0^2$ is
the Laplace operator on Euclidean 3-space $\mathbb{R}_3$,
which imposes the bound-state condition $w\leqslant \mu$.
Therefore,
similarly to other examples of spinning solutions with scalar fields
\cite{Radu:2008pp},
the presence of a mass
term in the action
is a necessary condition for the existence of localized configurations.

\begin{figure}[ht!]
\begin{center}
\includegraphics[height=.34\textwidth, angle =0 ]{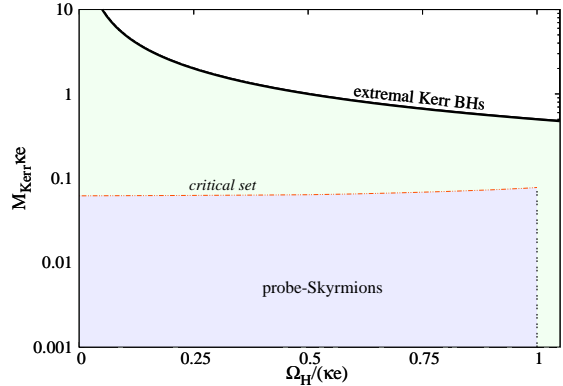}
%
%%\end{center}
\caption{
The domain of existence of Skerrmions in a Kerr BH mass $vs.$ horizon angular velocity diagram.
}
\label{domainKerr}
\end{center}
\end{figure}

Secondly, topological Skerrmions are found only for a limited set of Kerr backgrounds --  see Figure \ref{domainKerr},
with a maximal BH size defining the \textit{critical backgrounds}, similarly to what is observed in the static case, $cf.$ Fig.~\ref{Schw2}.
In the BH mass $vs.$ horizon angular velocity diagram,
the domain of existence of Skerrmions is bounded by a set of Schwarzschild BHs ($w=\Omega_H=0$),
the flat spacetime solutions ($M_{\rm Kerr}=0$ -- not seen in the plot),
the maximal frequency line $w=\Omega_H=\mu$,
and the set of critical backgrounds, wherein the solutions stop to exist, as in the static limit \cite{Luckock:1986tr}.
%%%%%%%%%%%%%%%%%%%%%%%%%%%%%%%%%%%%%%%%%%%%%%%%%%%%%%%%%%%%%%%%%%%%
%%%%%%%%%%%%%%%%%%%%%%%%%%%%%%%%%%%%%%%%%%%%%%%%%%%%%%%%%%%%%%%%%
%
%%%%%%%%%%%%%%%%%%%%%%%%%%%%%%%%%%%%%%%%%%%%%%%%%%%%%%%%%%%%%%%%
\begin{figure}[ht!]
%\lbfig{rhfar}
\begin{center}
\includegraphics[height=.34\textwidth, angle =0 ]{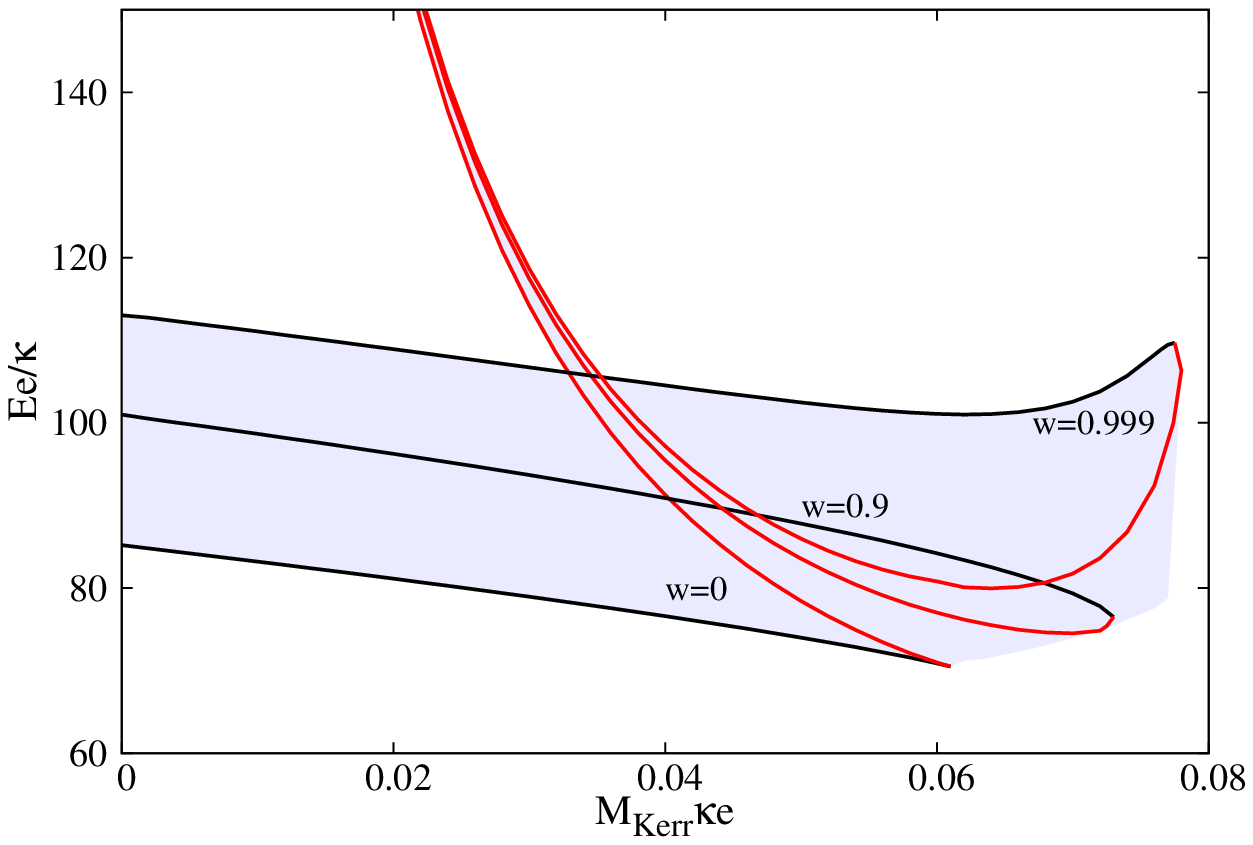}
\includegraphics[height=.34\textwidth, angle =0 ]{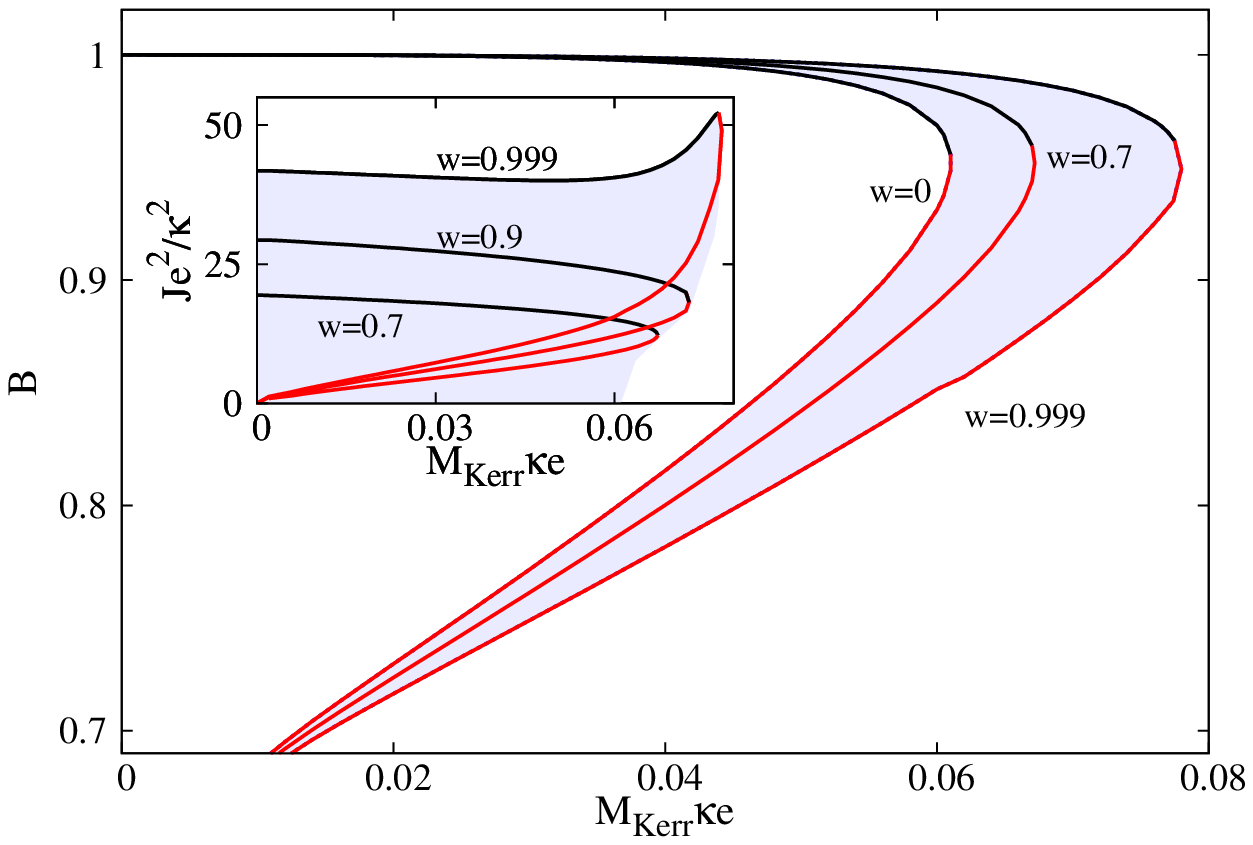}
\end{center}
\caption{
%The domain of existence in a Skerrmions mass-energy (left panel) \textit{or} baryon charge (right panel) \textit{or} angular momentum (inset) %$vs.$ the coupling parameter $\kappa e M_{\rm Kerr}$.
The Skerrmions mass-energy (left panel) \textit{or} baryon charge (right panel) \textit{or} angular momentum (inset)
ar shown $vs.$ the coupling parameter $\kappa e M_{\rm Kerr}$.
The shaded area corresponds to the explored domain of existence.
}
\label{probe}
\end{figure}
%%%%%%%%%%%%%%%%%%%%%%%%%%%%%%%%%%%%%%%%%%%%%%%%%%%%%%%%%%%%%%%%
%

Thirdly, we have also studied the domain of existence in a Skerrmions
mass-energy \textit{or} angular momentum \textit{or} baryon number $vs.$
the BH mass (or coupling constant) $M_{\rm Kerr}\kappa e$ diagram, as shown in Figure \ref{probe}. 
Figure \ref{probe} shows, for a given value of $w$, two branches
which
exists up to some maximal value  of the dimensionless parameter $M_{\rm Kerr}\kappa e$
(the maximal value increasing with $w$).
The limiting behaviour on these two branches is similar to the static case $(w=0)$.
The solutions on the fundamental branch are connected to flat space spinning Skyrmions,
while the limit of the excited branch is approached as
$\kappa \to 0$
and corresponds to the YM model on a fixed Schwarzschild background.
In particular, the angular momentum vanishes in this limit, and the above discussion for static
solutions is recovered. However, the second $\alpha\to 0$ limit is difficult to approach numerically

Fourthly,  the solutions tend to delocalize in the limit $w\to \mu$,
with the mass rapidly increasing,
while the numerics becomes challenging.

Finally, the value of the baryon charge $B$ decreases from $B=1$
(the flat space limit)
to a  minimal \textit{nonzero} value\footnote{Within the numerical errors,
this value corresponds to
the one found for $w=0$.}
which is approached along the second branch as $\alpha \to 0$.
%Thus, even in the presence of rotation, the BH cannot ``absorbe" the entire Skyrmion.
Thus, even in the presence
of rotation, the BH cannot ``absorbe" the entire Skyrmion, $i.e.$ the Skyrme field does not complete trivialise
outside the horizon.

In particular, as seen in Figure  \ref{domainKerr},
we do not find any indication for the occurrence of an existence line.
%
%

%%%%%%%%%%%%%%%%%%%%%%%%%%%%%%%%%%%%%%%%%%%%%%%%%%%%%%%%%%
%%%%%%%%%%%%%%%%%%%%%%%%%%%%%%%%%%%%%%%%%%%%%%%%%%%%%%%%%%
\section{Spinning BHs with Skyrme hair and non-topological Skerrmions}
\label{section5}
%%%%%%%%%%%%%%%%%%%%%%%%%%%%%%%%%%%%%%%%%%%%%%%%%%%%%%%%%%
%%%%%%%%%%%%%%%%%%%%%%%%%%%%%%%%%%%%%%%%%%%%%%%%%%%%%%%%%%

We shall now consider the coupled Einstein-Skyrme system, by supplementing (\ref{LS})
with the Einstein-Hilbert term. The action of the full model reads
\begin{eqnarray}
\label{action}
S=\int d^4 x  \sqrt{-g}
\left(\frac{R}{16 \pi G}+\mathcal{L}_S
\right) \ .
\end{eqnarray}

The spinning gravitating solutions are found for the same Skyrme field
Ansatz (\ref{U}) and line element (\ref{metric}); this time, however,
the metric functions $F_i$
and $W$
are different from those in the Kerr limit,
being found by solving numerically a boundary value problem. Thus,
in addition to the Skyrme equations
(\ref{Ueq}),
we also solve  the Einstein equations
\begin{eqnarray}
G_{\mu\nu}=8\pi G
 \,T_{\mu \nu},
\end{eqnarray}
with the stress-energy tensor
\begin{eqnarray}
\nonumber
T_{\mu\nu} &=&
-\frac{\kappa^2}{2}
\mbox{Tr}\!\left(K_\mu K_\nu-\frac{1}{2}g_{\mu \nu} K_\al K^\al \right)
- \frac{1}{8e^2}
\mbox{Tr}\!\left(
g^{\alpha \beta}\left[K_\mu,K_\alpha\right]\left[K_\nu,K_\beta\right]-
\frac{1}{4}g_{\mu \nu} \left[K_\alpha,K_\beta\right]\left[K^\alpha,K^\beta\right]
\right)
\\
\label{Tik}
&&
+ g_{\mu \nu}m_\pi^2\mbox{Tr}\left(\frac{U+U^\dagger}{2}- 
%
%\identity
\mathbf{1}
\right).
\end{eqnarray}
The boundary conditions in the matter sector
are similar to those employed in the probe limit,
while for the metric functions we impose $
F_i \big|_{r=\infty}=1,~~W \big|_{r=\infty} =0,~~
\partial_\theta F_i \big|_{\theta=0,\pi}=\partial_\theta W \big|_{\theta=0,\pi} =0,~~
\partial_r F_i \big|_{r=r_H}=0,~~W \big|_{r=r_H}=\Omega_H.$

The ADM  mass $M$ and the total angular momentum $J$ of the solutions
are read off from the asymptotics of the metric functions,
\begin{eqnarray}
\label{asym}
g_{tt} =-1+\frac{2GM}{r}+\dots \ , \qquad ~~g_{\varphi t}=-\frac{2GJ}{r}\sin^2\theta+\dots,
\end{eqnarray}
while the baryon charge $B$
is still given by (\ref{Qtop}).

Following
previous studies on hairy BHs~\cite{Delgado:2016zxv}, it is useful to introduce the parameter
\begin{eqnarray}
\label{p}
p \equiv \frac{M_H}{M}, ~~{\rm with}~~0\leqslant p\leqslant 1 \ ,
\end{eqnarray}
where $M_H$ is the horizon mass,
which measures the BH hairiness: $p=1$ for
Kerr BHs, for which $M_H=M$ corresponding to no matter fields outside horizon,
and $p=0$ for gravitating solitons, for which  $M_H=0$, since there is no  horizon.

Similarly to the test field case, one defines a dimensionless radial coordinate,
 frequency and a pion mass in terms of the input parameters $\kappa,e$.
However,
%as for static Einstein-Skyrme solutions
% \cite{Bizon:1992gb},
the presence of the Newton's constant $G$
in the theory's action (\ref{action})
introduces a new length scale.
Thus,
the gravitating model possesses a
 dimensionless coupling constant
\begin{eqnarray}
\alpha^2=4 \pi G  \kappa^2\ .
\end{eqnarray}
%which is the ratio of the two fundamental length scales of the problem.

Consequently, the  numerical problem has five input parameters.
The first two are geometrical: ${\bf i)}$ the event horizon angular velocity $\Omega_H$;
 ${\bf ii)}$ the event horizon radius $r_H$ in the metric ansatz (\ref{metric});
the next two characterizes the model:
 ${\bf iii)}$ the coupling constant $\alpha$ and
 ${\bf iv)}$ the (scaled) pion mass  $\mu$.
The last input parameter\footnote{As we shall see, the Skyrme field frequency
(which is another input parameter)
 is fixed by $m$ and $\Omega_H$ via the synchronization condition (\ref{synch}).}
is
 ${\bf v)}$ the winding number $m$.
%The first two parameters are geometric quantities, while the next two characterizes the theory.
All displayed numerical results
were found for a choice of  (scaled) pion mass $\mu=1$ and winding number $m=1$.

%%%%%%%%%%%%%%%%%%%%%%%%%%%%%%%%%%%%%%%%%%%%%%%%%%%%%%%%%%
\subsection{The topological sector}
%%%%%%%%%%%%%%%%%%%%%%%%%%%%%%%%%%%%%%%%%%%%%%%%%%%%%%%%%%

The solutions with $B\neq 0$ define  the topological sector of spinning BHs with Skyrme hair.
The emerging picture in the backreacting case
 is more complicated than in the probe limit,
with new qualitative features,
which we have only started to unveil.
Its basic features can be summarize as follows.

All known static BHs with Skyrme hair
\cite{Bizon:1992gb}
possess rotating generalizations,
which are found by
increasing gradually the
event horizon velocity $\Omega_H$.
Alternatively, a different route to construct the spinning BH solutions with topological Skyrme hair
is to start with the  spinning gravitating solitons in
 \cite{Ioannidou:2006nn} (with a given set of relevant parameters)
and to slowly increase the
horizon size via the parameter $r_H$.

Similarly to the test field limit,
the synchronous condition
(\ref{synch}) must hold,
with $w=m\Omega_H$.
 Then, the spinning hairy BHs are found for all range $0\leqslant w < \mu$,
 where $w,\mu$ are the scaled Skyrme field frequency and mass.

Rotating BHs with Skyrme hair inherit
from the solitonic limit \cite{Ioannidou:2006nn}
a  complex branch
structure in terms of $\alpha$, a full analysis of which is beyond the scope of this work.
Some basic aspects are as follows. For a given value of the (scaled) frequency $w$
and a given horizon size,
one set of solutions form  the fundamental branch.
The limit  $\alpha \to 0$ corresponds to the test field Skerrmions.
One finds also a branch of
excited solutions
ending in the (static) Einstein-YM system as
$\alpha \to 0$ (which corresponds along this branch to $\kappa \to 0$).
This is the behaviour found already in the static limit \cite{Bizon:1992gb};
the arguments therein also applying for $w\neq 0$.
%
%These two branches possess a static limit, in which case the spherically symmetric
However, in the rotating case there exist (at least) two more branches of excited solutions within the full ansatz (\ref{U}),
which  do not possess a well-defined static limit,
nor a Kerr-probe one.

\begin{figure}[ht!]
%\lbfig{rhfar}
\begin{center}
\includegraphics[height=.34\textwidth, angle =0 ]{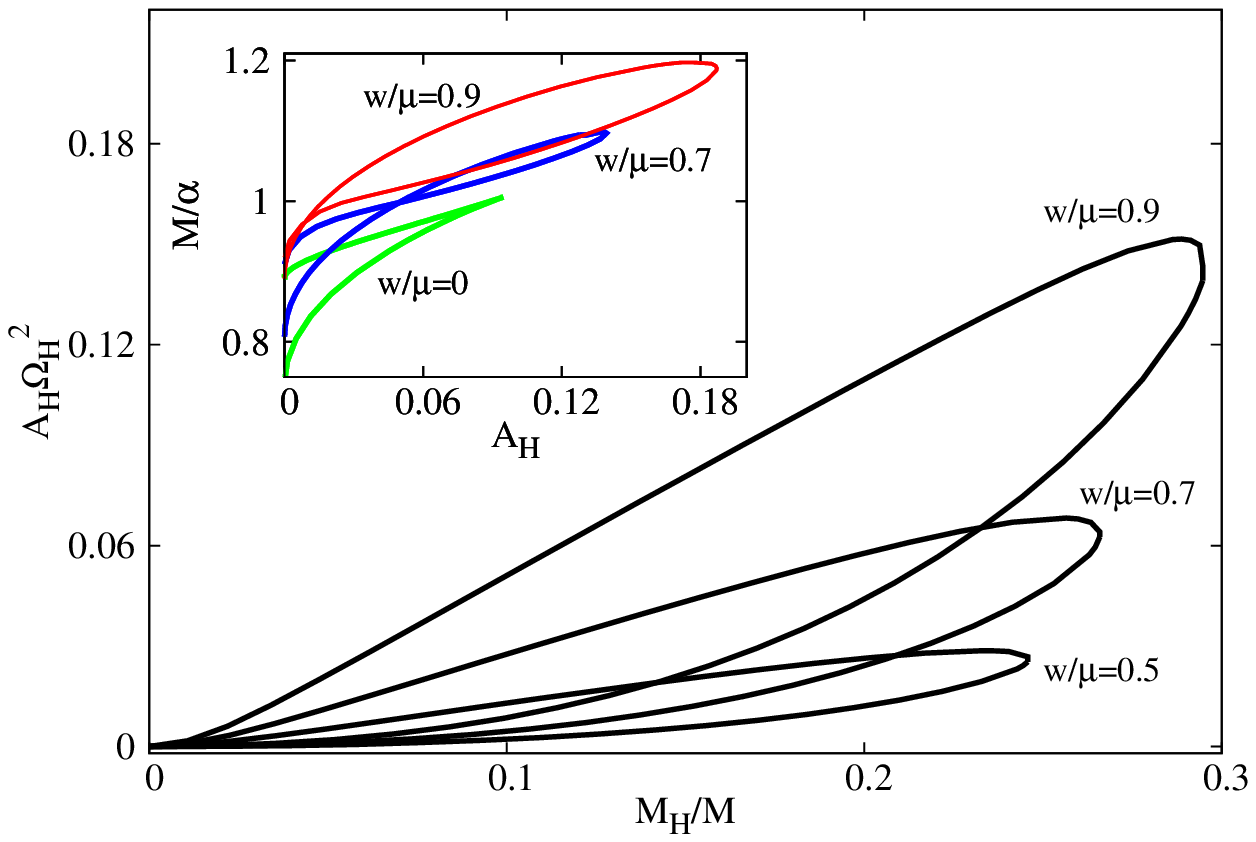}
\includegraphics[height=.34\textwidth, angle =0 ]{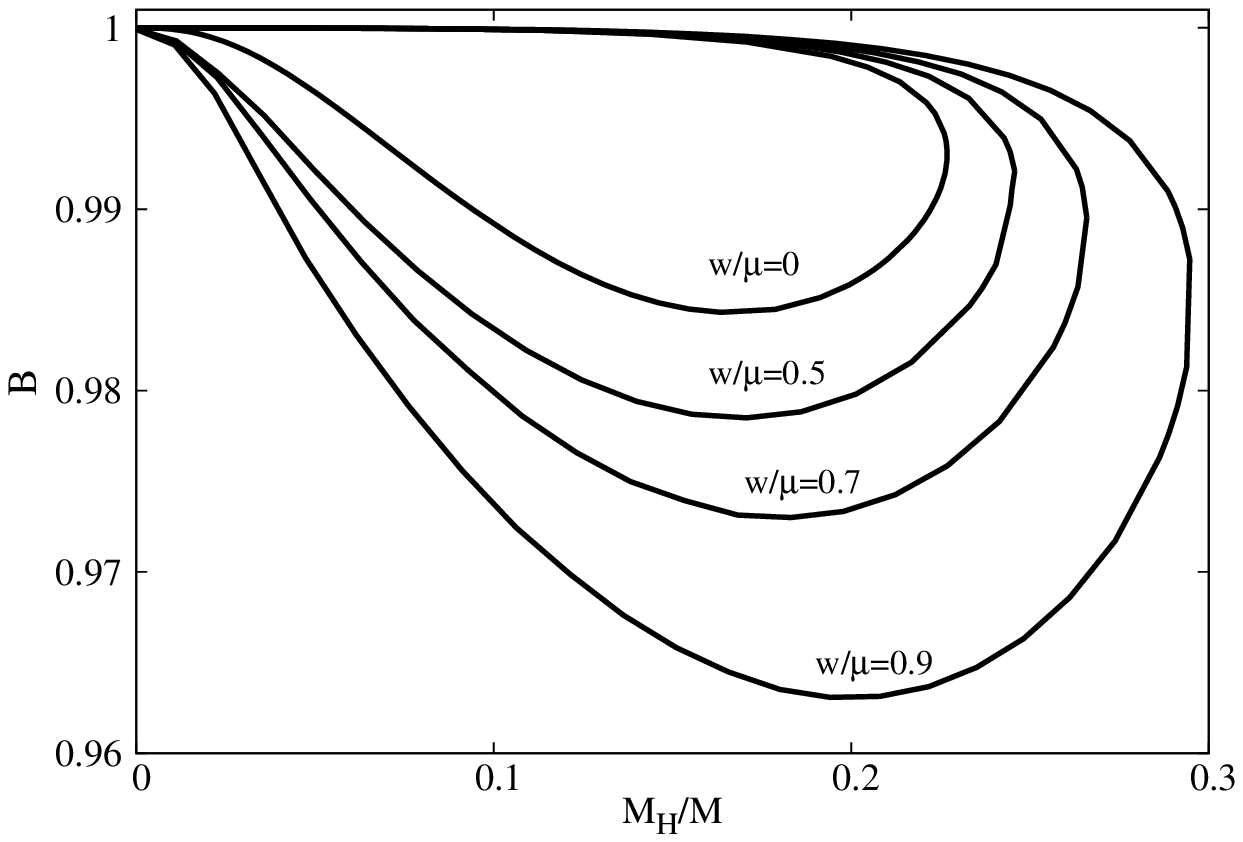}
\end{center}
\caption{
The dimensionless parameter $A_H \Omega_H^2$ (left panel)
and the baryon charge $B$ (right panel) as functions of the hairiness parameter $p=M_H/M$
for BH solutions with the coupling constant
$\alpha=0.12$.
}
\label{grav1}
\end{figure}

Some basic properties of solutions
with a generic value of $\alpha$
as a function of the hairiness parameter $p=M_H/M$
can be seen in Figure \ref{grav1}.
One can notice $e.g.$ the existence of
a two branch structure of solutions in terms of $A_H$
(or, equivalently the horizon radius $r_H$), with
a maximal value of the horizon area.
Again, the properties of the static,
spherically symmetric  case ($w=0$) appear to be generic.
The lower branch emerges from the (fundamental branch of)
gravitating Skyrmions in  \cite{Ioannidou:2006nn}
with a given $\alpha$.
There is also a second branch of (excited) solutions
which coalesces with the first one for a maximal value of the hairiness parameter $p$.
However,  in this case
the limit $A_H\to 0$ is approached as
$\kappa \to 0$.
%in which case the  solutions of the (spherically symmetric) Einstein--Yang-Mills (EYM) model are recovered.
 %
Then,
after a suitable rescaling
the fundamental
spherically symmetric Bartnik-McKinnon particle-like solution  of the Einstein-YM system
(as reviewed $e.g.$ in \cite{Volkov:1998cc})
 are recovered.
%

%%%%%%%%%%%%%%%%%%%%%%%%%%%%%%%%%%%%%%%%%%%%%%%%%%%%%%%%%%%%%%%%
%%%%%%%%%%%%%%%%%%%%%%%%%%%%%%%%%%%%%%%%%%%%%%%%%%%%%%%%%%%%%%%%
%

Most importantly in the context of this work,
no indication is found for the presence of an existence line.
That is, for all  branches,
 the Skyrme field never trivializes ($p<1$) and
the baryon charge $B$ is always strictly positive, see Figure \ref{grav1} (right panel).
%its minimal value decreasing with the parameter $w/\mu$.
%

%%%%%%%%%%%%%%%%%%%%%%%%%%%%%%%%%%%%%%%%%%%%%%%%%%%%%%%%%%
\subsection{The non-topological sector}
%%%%%%%%%%%%%%%%%%%%%%%%%%%%%%%%%%%%%%%%%%%%%%%%%%%%%%%%%%
The $B\neq 0$ solutions
discussed
in the previous subsection
were found starting with (topologically nontrivial) Skyrme solitons.
Disconnected branches of configurations with a vanishing baryon charge $B$ also exist. These solutions are found for a consistent truncation
of the Skyrme model with
\begin{eqnarray}
\label{Us}
 \pi_1+i \pi_2=\phi_1(r,\theta)e^{i(m\varphi-w t)},~~
 \pi_3 \equiv 0, ~~\sigma=\phi_3(r,\theta), ~~~{\rm with}~~~ \phi_1^2+\phi_3^2=1 \ .
\end{eqnarray}
The existence of this Ansatz has been noticed already in
\cite{Ioannidou:2006nn},
the corresponding solitons being dubbed
\textit{pion clouds}.
Since
$\pi_3 \equiv 0$,
the topological charge vanishes identically in this case, $B=0$. With this truncation,
the Skyrme model
collapses to the $O(3)$-Fadeev-Skyrme sigma-model \cite{Radu:2008pp}.

The small field linearization of the model (\ref{Us})
around the vacuum state $\sigma=1$
results in
the Lagrangean (\ref{s2})
with $\chi=0$ and
\begin{eqnarray}
\psi=\phi_1(r,\theta)e^{i(m\varphi-w t)}.
\end{eqnarray}
As discussed above, $\psi$ posses bound state solutions for
a particular one-dimensional set of
Kerr backgrounds forming the existence line.
In what follows, we show that these `scalar clouds'
can be promoted to
fully non-linear  solutions of the
Einstein-Skyrme model,
provided one uses the truncated Skyrme Ansatz  (\ref{Us}).

%%%%%%%%%%%%%%%%%%%%%%%%%%%%%%%%%%%%%%%%%%%%%%%%%%%%%%%%%%%%%%%%
\begin{figure}[ht!]
%\lbfig{rhfar}
\begin{center}
\includegraphics[height=.34\textwidth, angle =0 ]{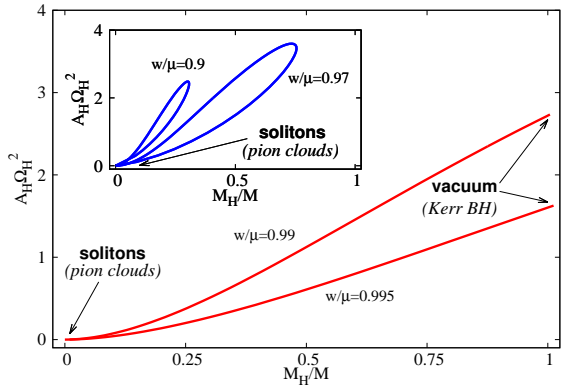}
\end{center}
\caption{
The dimensionless parameter $A_H \Omega_H^2$ as a function of the hairiness parameter $p=M_H/M$
for spinning BHs with non-topological Skyrme hair,  for a coupling constant $\alpha=10$.
}
\label{grav2}
\end{figure}
%%%%%%%%%%%%%%%%%%%%%%%%%%%%%%%%%%%%%%%%%%%%%%%%%%%%%%%%%%%%%%%%

Following the same approach (and the same boundary conditions) as for the topological sector, we have studied BH
solutions with non-topological Skyrme hair, within this particular model.
%The basic properties of the solutions
%can be summarized as follows.
%
Rather surprisingly, they are akin to the Kerr BHs with scalar hair found in~\cite{Herdeiro:2014goa,Herdeiro:2015gia},  sharing
most of their features, not to solutions of the full Skyrme model. For example,
in strong contrast to the Skyrme case,
their existence requires an Einstein term in the action
(this holds both for BHs and solitons).
As such, they do not possess a flat space limit;
moreover, no other solutions on
a fixed Kerr background is found apart from the scalar clouds along the existence line.

Figure \ref{grav2} shows the BH horizon area for these hairy BHs (normalised as $A_H\Omega_H^2$) $vs.$ the hairiness parameter $p=M_H/M$.
One observes, for instance, that in the limit of a vanishing horizon mass (a $p=0$ hairiness parameter) all solutions reduce to the solitonic
`pion clouds' discussed in
\cite{Ioannidou:2006nn}.
These are a more complicated version of the well-known spinning boson stars~\cite{Yoshida:1997qf}.
In contrast to the case of solutions of the full model ($B\neq 0$),
one notices the existence of branches of
solutions  which end on vacuum Kerr BHs ($p=M_H/M=1$).
This feature occurs for a range of field frequencies $0.95 < w/\mu <1$.
As such those sets of solutions
interpolate between solitons and  Kerr BHs on the existence line, and they can be considered as nonlinear realizations of the stationary scalar clouds first discussed in~\cite{Hod:2012px} within the  Einstein-Skyrme system.
Branches of BH solutions interpolating between
two different solitons with the same field frequency exist as well for small enough values of $w$
($cf.$ the inset in Figure \ref{grav2}). In this case $p$ is always smaller than one.

Finally,
the spinning BHs with non-topological Skyrme hair do not possess a static limit, $w=\Omega_H=0$.
Below a minimal value of the field frequency no hairy BHs are found.
%

%%%%%%%%%%%%%%%%%%%%%%%%%%%%%%%%%%%%%%%%%%%%%%%%%%%%%%%%%%%%%%%%
%%%%%%%%%%%%%%%%%%%%%%%%%%%%%%%%%%%%%%%%%%%%%%%%%%%%%%%%%%%%%%%%
\section{Conclusions}
\label{section6}
%%%%%%%%%%%%%%%%%%%%%%%%%%%%%%%%%%%%%%%%%%%%%%%%%%%%%%%%%%%%%%%%
%%%%%%%%%%%%%%%%%%%%%%%%%%%%%%%%%%%%%%%%%%%%%%%%%%%%%%%%%%%%%%%%
Static, spherically symmetric BHs with Skyrme hair are a pioneering counter-example to the ``no-hair" conjecture, at least in its strongest version. But up to now, the effect of BH rotation upon this ``hair" had not been considered.  In this work, we have shown the existence of solutions of the Skyrme model on the Kerr BH background, as test field solutions -- Skerrmions -- , and also constructed four dimensional spinning BHs with Skyrme hair, which may be regarded as back reacting Skerrmions  (see~{\cite{Brihaye:2017wqa} for higher dimensional spinning BHs with Skyrme hair, where a technical advantage occurs).

For the existence of spinning BHs with Skyrme  ``hair" the synchronisation condition~\eqref{synch} is crucial.\footnote{The ``hair" in the static, spherically symmetric case can be faced as obeying a trivial version of this condition.} We have found there are two sectors of such hairy solutions. The first one is a topological sector qualitatively similar to that found in the static case. The second one is a non-topological sector, qualitatively different from the solutions found in the static case, and more akin to Kerr BHs with synchronised scalar hair~\cite{Herdeiro:2014goa,Herdeiro:2015gia}. The exploration performed in this work has really just scratched the surface of the rich structure of solutions of the Skyrme-Einstein model, which certainly merits to be analysed in greater depth.

%%%%%%%%%%%%%%%%%%%%%%%%%%%%%%%%%%%%%%%%%%%%%%%%%%%%%%%%%%%%%%%%%%%%%
\section*{Acknowledgements}
%\noindent{\bf{\em Acknowledgements.}}
%%%%%%%%%%%%%%%%%%%%%%%%%%%%%%%%%%%%%%%%%%%%%%%%%%%%%%%%%%%%%%%%%%%%%
This work has been supported by the FCT (Portugal) IF programme, by the
FCT grant PTDC/FISOUT/28407/2017, by the CIDMA (FCT) strategic project
UID/MAT/04106/2013, by the CENTRA (FCT) strategic project
UID/FIS/00099/2013 and
by  the  European  Union's  Horizon  2020  research  and  innovation  (RISE) programmes H2020-MSCA-RISE-2015
Grant No.~StronGrHEP-690904 and H2020-MSCA-RISE-2017 Grant No.~FunFiCO-777740. The authors would like to acknowledge
networking support by the
COST Action CA16104. Y.S. gratefully
acknowledges support from the Ministry of Education and Science
of Russian Federation, project No 3.1386.2017.
The parallel computations were performed on the cluster HIBRILIT at LIT, JINR, Dubna and Blafis at Aveiro University.

%%%%%%%%%%%%%%%%%%%%%%%%%%%%%%%%%%%%%%%%%%%%%%%%%%%%%%%%%%%%%%%%%%%%%%%%%%%%%%%%%%%%  
 \begin{small}

%%%%%%%%%%%%%%%%%%%%%%%%%%%%%%%%%%%%%%%%%%%%%%%%%%%%%%%%%%%%

%%%%%%%%%%%%%%%%%%%%%%%%%%%%%%%%%%%%%%%%%%%%%%%%%%%%%%%%%%%%

 \end{small}


\begin{thebibliography}{99}
 %%%%%%%%%%%%%%%%%%%%%%%%%%%%%%%%%%%%%%%%%%%%%%%%%%%%%%%%%%%%
 %\cite{Ruffini:1971bza}
\bibitem{Ruffini:1971bza}
  R.~Ruffini and J.~A.~Wheeler,
  ``Introducing the black hole,''
  Phys.\ Today {\bf 24} (1971) no.1,  30.
  %doi:10.1063/1.3022513
  %%CITATION = doi:10.1063/1.3022513;%%
  %216 citations counted in INSPIRE as of 10 Oct 2017
%%%%%%%%%%%%%%%%%%%%%%%%%%%%%%%%%%%%%%%%%%%%%%%%%%%%%%%%%%%%%%%%%%%%% 
%%%%%%%%%%%%%%
%\cite{Bekenstein:1996pn}
\bibitem{Bekenstein:1996pn}
  J.~D.~Bekenstein,
   {\it Black hole hair: 25 - years after,}
%  in {\it Moscow 1996, 2nd International A.D. Sakharov Conference on physics}, pp. 216-219,
  [gr-qc/9605059].
  %%CITATION = GR-QC/9605059;%%
  %54 citations counted in INSPIRE as of 04 Jun 2013
    %%%%%%%%%%%%%%%%%%%%%%%%%%%%%%%%%%%%%%%%%%%%%%%%%%%%%%%%%%%%%%%%%%%%%%%%%%%%%%
%\cite{Herdeiro:2015waa}
\bibitem{Herdeiro:2015waa}
  C.~A.~R.~Herdeiro and E.~Radu,
  ``Asymptotically flat black holes with scalar hair: a review,''
  Int.\ J.\ Mod.\ Phys.\ D {\bf 24} (2015) no.09,  1542014
 % doi:10.1142/S0218271815420146
  [arXiv:1504.08209 [gr-qc]].
  %%CITATION = doi:10.1142/S0218271815420146;%%
  %117 citations counted in INSPIRE as of 30 Sep 2017
%%%%%%%%%%%%%%%%%%%%%%%%%%%%%%%%%%%%%%%%%%%%%%%%%%%%%%%%%%%%%%%%%%%
%\cite{Sotiriou:2015pka}
\bibitem{Sotiriou:2015pka}
  T.~P.~Sotiriou,
  ``Black holes and scalar fields,''
  Class.\ Quant.\ Grav.\  {\bf 32} (2015) no.21,  214002
  %doi:10.1088/0264-9381/32/21/214002
  [arXiv:1505.00248 [gr-qc]].
  %%CITATION = doi:10.1088/0264-9381/32/21/214002;%%
  %38 citations counted in INSPIRE as of 16 Jul 2018
%%%%%%%%%%%%%%%%%%%%%%%%%%%%%%%%%%%%%%%%%%%%%%%%%%%%%%%%%%%%%%%%%%%
%%%%%%%%%%%%%%%%%%%%%%%%%%%%%%%%%%%%%%%%%%%%%%%%%%%%%%%%%%%%%%%%%%%
%\cite{Volkov:2016ehx}
\bibitem{Volkov:2016ehx}
  M.~S.~Volkov,
  ``Hairy black holes in the XX-th and XXI-st centuries,''
  %doi:10.1142/9789813226609_0184
  arXiv:1601.08230 [gr-qc].
  %%CITATION = doi:10.1142/9789813226609_0184;%%
  %39 citations counted in INSPIRE as of 16 Jul 2018
%%%%%%%%%%%%%%%%%%%%%%%%%%%%%%%%%%%%%%%%%%%%%%%%%%%%%%%%%%%%%%%%%%%
%%%%%%%%%%%%%%%%%%%%%%%%%%%%%%%%%%%%%%%%%%%%%%%%%%%%%%%%%%%%%%%%%%%
%\cite{Skyrme:1961vq}\cite{Skyrme:1962vh}
\bibitem{Skyrme:1961vq}
  T.~H.~R.~Skyrme,
  ``A nonlinear field theory,''
  Proc.\ Roy.\ Soc.\ Lond.\ A {\bf 260} (1961) 127.
 % doi:10.1098/rspa.1961.0018
  %%CITATION = doi:10.1098/rspa.1961.0018;%%
  %1962 citations counted in INSPIRE as of 21 Sep 2017
%%%%%%%%%%%%%%%%%%%%%%%%%%%%%%%%%%%%%%%%%%%%%%%%%%%%%%%%%%%%%%%%%%%%%%%%%%%%%%%%%%%%%%%%%%%%%%%%%%%%%%%%%
%\cite{Skyrme:1962vh}
\bibitem{Skyrme:1962vh}
  T.~H.~R.~Skyrme,
 ``A unified field theory of mesons and baryons,''
  Nucl.\ Phys.\  {\bf 31} (1962) 556.
  %%CITATION = NUPHA,31,556;%%
    %%%%%%%%%%%%%%%%%%%%%%%%%%%%%%%%%%%%%%%%%%%%%%%%%%%%%%%%%%%%%%%%%%%%%%%%%%%%%%%%%%%%%%%%%%%%%%%%%%%%%%%%%
%\cite{Witten:1983tw}
\bibitem{Witten:1983tw}
  E.~Witten,
  ``Global aspects of current algebra,''
  Nucl.\ Phys.\ B {\bf 223} (1983) 422;
%  doi:10.1016/0550-3213(83)90063-9
  %%CITATION = doi:10.1016/0550-3213(83)90063-9;%%
  %2444 citations counted in INSPIRE as of 24 Sep 2017
    \\
    %\cite{Witten:1983tx}
%\bibitem{Witten:1983tx}
  E.~Witten,
  ``Current algebra, baryons, and quark confinement,''
  Nucl.\ Phys.\ B {\bf 223} (1983) 433.
 % doi:10.1016/0550-3213(83)90064-0
  %%CITATION = doi:10.1016/0550-3213(83)90064-0;%%
  %1261 citations counted in INSPIRE as of 24 Sep 2017
%%%%%%%%%%%%%%%%%%%%%%%%%%%%%%%%%%%%%%%%%%%%%%%%%%%%%%%%%%%%%%%%%%%%%%%%%%%%%%%%%%%%%%%%%%%%%%%%%%%%%%%%%
%\cite{Callan:1983nx}
\bibitem{Callan:1983nx}
  C.~G.~Callan, Jr. and E.~Witten,
  ``Monopole catalysis of Skyrmion decay,''
  Nucl.\ Phys.\ B {\bf 239} (1984) 161.
 % doi:10.1016/0550-3213(84)90088-9
  %%CITATION = doi:10.1016/0550-3213(84)90088-9;%%
  %80 citations counted in INSPIRE as of 24 Sep 2017
 %%%%%%%%%%%%%%%%%%%%%%%%%%%%%%%%%%%%%%%%%%%%%%%%%%%%%%%%%%%%%%%%%%
%%%%%%%%%%%%%%%%%%%%%%%%%%%%%%%%%%%%%%%%%%%%%%%%%%%%%%%%%%%%%%%%%%
%\cite{Luckock:1986tr}
\bibitem{Luckock:1986tr}
  H.~Luckock and I.~Moss,
  ``Black holes have Skyrmion hair,''
  Phys.\ Lett.\ B {\bf 176} (1986) 341.
%  doi:10.1016/0370-2693(86)90175-9
  %%CITATION = doi:10.1016/0370-2693(86)90175-9;%%
  %82 citations counted in INSPIRE as of 04 Nov 2017
  %%%%%%%%%%%%%%%%%%%%%%%%%%%%%%%%%%%%%%%%%%%%%%%%%%%%%%%%%%%%%%%%%%%
%
%\cite{Luckock}
\bibitem{Luckock}
H. Luckock, {\it Black hole skyrmions}, in H.J. De Vega and N. Sanches, editors,
{\it String Theory, Quantum Cosmology and Quantum Gravity, Integrable and
Conformal Integrable Theories} pp. 455, World Scientific, 1987.
%
%%%%%%%%%%%%%%%%%%%%%%%%%%%%%%%%%%%%%%%%%%%%%%%%%%%%%%%%%%%%%%%%%%%%%%%%
    %\cite{Droz:1991cx}
\bibitem{Droz:1991cx}
  S.~Droz, M.~Heusler and N.~Straumann,
  ``New black hole solutions with hair,''
  Phys.\ Lett.\ B {\bf 268} (1991) 371.
  %doi:10.1016/0370-2693(91)91592-J
  %%CITATION = doi:10.1016/0370-2693(91)91592-J;%%
  %134 citations counted in INSPIRE as of 30 Oct 2016
%%%%%%%%%%%%%%%%%%%%%%%%%%%%%%%%%%%%%%%%%%%%%%%%%%%%%%%%%%%%%%%%%%%%%%%%%%%%%%%%%%%%%%%%%%%%%%%%%%%%%%%%%
%\cite{Adam:2016vzf}
\bibitem{Adam:2016vzf}
  C.~Adam, O.~Kichakova, Y.~Shnir and A.~Wereszczynski,
  ``Hairy black holes in the general Skyrme model,''
  Phys.\ Rev.\ D {\bf 94} (2016) no.2,  024060
 % doi:10.1103/PhysRevD.94.024060
  [arXiv:1605.07625 [hep-th]].
  %%CITATION = doi:10.1103/PhysRevD.94.024060;%%
  %10 citations counted in INSPIRE as of 08 Oct 2017
%%%%%%%%%%%%%%%%%%%%%%%%%%%%%%%%%%%%%%%%%%%%%%%%%%%%%%%%%%%%%%%%%%%%%%%%
%\cite{Dvali:2016mur}
\bibitem{Dvali:2016mur}
  G.~Dvali and A.~GuÃŸmann,
  ``Skyrmion black hole hair: conservation of baryon number
    by black holes and observable manifestations,''
  Nucl.\ Phys.\ B {\bf 913} (2016) 1001
 % doi:10.1016/j.nuclphysb.2016.10.017
  [arXiv:1605.00543 [hep-th]].
  %%CITATION = doi:10.1016/j.nuclphysb.2016.10.017;%%
  %14 citations counted in INSPIRE as of 08 Oct 2017
%%%%%%%%%%%%%%%%%%%%%%%%%%%%%%%%%%%%%%%%%%%%%%%%%%%%%%%%%%%%%%%%%%%%%%%%
    %\cite{Gudnason:2016kuu}
\bibitem{Gudnason:2016kuu}
  S.~B.~Gudnason, M.~Nitta and N.~Sawado,
  ``Black hole Skyrmion in a generalized Skyrme model,''
  JHEP {\bf 1609} (2016) 055
 % doi:10.1007/JHEP09(2016)055
  [arXiv:1605.07954 [hep-th]].
  %%CITATION = doi:10.1007/JHEP09(2016)055;%%
  %10 citations counted in INSPIRE as of 08 Oct 2017
    %%%%%%%%%%%%%%%%%%%%%%%%%%%%%%%%%%%%%%%%%%%%%%%%%%%%%%%%%%%%%%%%%%%%%%%%%%%%%%%%%%%%%%%%%%%%%%%%%%%%%%%%%
%%%%%%%%%%%%%%%%%%%%%%%%%%%%%%%%%%%%%%%%%%%%%%%%%%%%%%%%%%%%%%%%%%%%%%%%%%%%%%%%%%%%%%%%%%%%%%%%%%%%%%%%%
%\cite{Glendenning:1988qy}
\bibitem{Glendenning:1988qy}
  N.~K.~Glendenning, T.~Kodama and F.~R.~Klinkhamer,
  ``Skyrme topological soliton coupled to gravity,''
  Phys.\ Rev.\ D {\bf 38} (1988) 3226.
  %doi:10.1103/PhysRevD.38.3226
  %%CITATION = doi:10.1103/PhysRevD.38.3226;%%
  %30 citations counted in INSPIRE as of 21 Sep 2017
%%%%%%%%%%%%%%%%%%%%%%%%%%%%%%%%%%%%%%%%%%%%%%%%%%%%%%%%%%%%%%%%%%%%%%%%
%\cite{Bizon:1992gb}
\bibitem{Bizon:1992gb}
  P.~Bizon and T.~Chmaj,
  ``Gravitating skyrmions,''
  Phys.\ Lett.\ B {\bf 297} (1992) 55.
  %doi:10.1016/0370-2693(92)91069-L
  %%CITATION = doi:10.1016/0370-2693(92)91069-L;%%
  %99 citations counted in INSPIRE as of 21 Sep 2017
%%%%%%%%%%%%%%%%%%%%%%%%%%%%%%%%%%%%%%%%%%%%%%%%%%%%%%%%%%%%%%%%%%%%%%%%
%\cite{Heusler:1991xx}
\bibitem{Heusler:1991xx}
  M.~Heusler, S.~Droz and N.~Straumann,
  ``Stability analysis of self-gravitating skyrmions,''
  Phys.\ Lett.\ B {\bf 271} (1991) 61.
  %doi:10.1016/0370-2693(91)91278-4
  %%CITATION = doi:10.1016/0370-2693(91)91278-4;%%
  %72 citations counted in INSPIRE as of 30 Oct 2016
%%%%%%%%%%%%%%%%%%%%%%%%%%%%%%%%%%%%%%%%%%%%%%%%%%%%%%%%%%%%%%%%%%%%%%%%%%%%%%%%%%%%%%%%%%%%%%%%%%%%%%%%%
%\cite{Heusler:1993ci}
\bibitem{Heusler:1993ci}
  M.~Heusler, N.~Straumann and Z.~h.~Zhou,
  ``Selfgravitating solutions of the Skyrme model and their stability,''
  Helv.\ Phys.\ Acta {\bf 66} (1993) 614.
  %%CITATION = HPACA,66,614;%%
  %33 citations counted in INSPIRE as of 08 Oct 2017
%%%%%%%%%%%%%%%%%%%%%%%%%%%%%%%%%%%%%%%%%%%%%%%%%%%%%%%%%%%%%%%%%%%%%%%%%%%%%%%%%%%%%%%%%%%%%%%%%%%%%%%%%
%\cite{Battye:2005nx}
\bibitem{Battye:2005nx}
  R.~A.~Battye, S.~Krusch and P.~M.~Sutcliffe,
  ``Spinning skyrmions and the skyrme parameters,''
  Phys.\ Lett.\ B {\bf 626} (2005) 120
  %doi:10.1016/j.physletb.2005.08.097
  [hep-th/0507279].
  %%CITATION = doi:10.1016/j.physletb.2005.08.097;%%
  %53 citations counted in INSPIRE as of 30 Oct 2016
%%%%%%%%%%%%%%%%%%%%%%%%%%%%%%%%%%%%%%%%%%%%%%%%%%%%%%%%%%%%%%%%%%%%%%%%%%%%%%%%%%%
%\cite{Ioannidou:2006nn}
\bibitem{Ioannidou:2006nn}
  T.~Ioannidou, B.~Kleihaus and J.~Kunz,
  ``Spinning gravitating skyrmions,''
  Phys.\ Lett.\ B {\bf 643} (2006) 213
 % doi:10.1016/j.physletb.2006.10.055
  [gr-qc/0608110].
  %%CITATION = doi:10.1016/j.physletb.2006.10.055;%%
  %11 citations counted in INSPIRE as of 30 Oct 2016
    %%%%%%%%%%%%%%%%%%%%%%%%%%%%%%%%%%%%%%%%%%%%%%%%%%%%%%%%%%%%%%%%%%
%\cite{Perapechka:2017bsb}
\bibitem{Perapechka:2017bsb}
  I.~Perapechka and Y.~Shnir,
  ``Spinning gravitating Skyrmions in a generalized Einstein-Skyrme model,''
  Phys.\ Rev.\ D {\bf 96} (2017) no.12,  125006
 % doi:10.1103/PhysRevD.96.125006
  [arXiv:1710.06334 [hep-th]].
  %%CITATION = doi:10.1103/PhysRevD.96.125006;%%
  %1 citations counted in INSPIRE as of 02 May 2018
	

%%%%%%%%%%%%%%%%%%%%%%%%%%%%%%%%%%%%%%%%%%%%%%%%%%%%%%%%%%%%%%%%%%%%%%%%%%%%%%%%%%
%\cite{Hod:2012px}
\bibitem{Hod:2012px}
  S.~Hod,
  ``Stationary scalar clouds around rotating black holes,''
  Phys.\ Rev.\ D {\bf 86} (2012) 104026
   Erratum: [Phys.\ Rev.\ D {\bf 86} (2012) 129902]
  %doi:10.1103/PhysRevD.86.129902, 10.1103/PhysRevD.86.104026
  [arXiv:1211.3202 [gr-qc]].
  %%CITATION = doi:10.1103/PhysRevD.86.129902, 10.1103/PhysRevD.86.104026;%%
  %111 citations counted in INSPIRE as of 26 Jun 2018
 %%%%%%%%%%%%%%%%%%%%%%%%%%%%%%%%%%%%%%%%%%%%%%%%%%%%%%%%%%%	
%%%%%%%%%%%%%%%%%%%%%%%%%%%%%%%%%%%%%%%%%%%%%%%%%%%%%%%%%%%%
%\cite{Herdeiro:2014goa}
\bibitem{Herdeiro:2014goa}
  C.~A.~R.~Herdeiro and E.~Radu,
  ``Kerr black holes with scalar hair,''
  Phys.\ Rev.\ Lett.\  {\bf 112} (2014) 221101
  %doi:10.1103/PhysRevLett.112.221101
  [arXiv:1403.2757 [gr-qc]].
  %%CITATION = doi:10.1103/PhysRevLett.112.221101;%%
  %228 citations counted in INSPIRE as of 28 Sep 2017
%%%%%%%%%%%%%%%%%%%%%%%%%%%%%%%%%%%%%%%%%%%%%%%%%%%%%%%%%%%%%%%%%%%%%%%%%%%%%%%%%%%
%\cite{Herdeiro:2015gia}
\bibitem{Herdeiro:2015gia}
  C.~Herdeiro and E.~Radu,
  ``Construction and physical properties of Kerr black holes with scalar hair,''
  Class.\ Quant.\ Grav.\  {\bf 32} (2015) no.14,  144001
  %doi:10.1088/0264-9381/32/14/144001
  [arXiv:1501.04319 [gr-qc]].
  %%CITATION = doi:10.1088/0264-9381/32/14/144001;%%
  %105 citations counted in INSPIRE as of 07 Oct 2017
%%%%%%%%%%%%%%%%%%%%%%%%%%%%%%%%%%%%%%%%%%%%%%%%%%%%%%%%%%%%%%%%%%%%%%%%%%%%%%%%%%%
%\cite{Brihaye:2014nba}
\bibitem{Brihaye:2014nba}
  Y.~Brihaye, C.~Herdeiro and E.~Radu,
  ``Myers-Perry black holes with scalar hair and a mass gap,''
  Phys.\ Lett.\ B {\bf 739} (2014) 1
  %doi:10.1016/j.physletb.2014.10.019
  [arXiv:1408.5581 [gr-qc]].
  %%CITATION = doi:10.1016/j.physletb.2014.10.019;%%
  %61 citations counted in INSPIRE as of 16 Jul 2018
  %%%%%%%%%%%%%%%%%%%%%%%%%%%%%%%%%%%%%%%%%%%%%%%%%%%%%%%%%%%%%%%%%%%%%%%%%%%%%%%%%%%
%\cite{Kleihaus:2015iea}
\bibitem{Kleihaus:2015iea}
  B.~Kleihaus, J.~Kunz and S.~Yazadjiev,
  ``Scalarized hairy black holes,''
  Phys.\ Lett.\ B {\bf 744} (2015) 406
  %doi:10.1016/j.physletb.2015.04.014
  [arXiv:1503.01672 [gr-qc]].
  %%CITATION = doi:10.1016/j.physletb.2015.04.014;%%
  %46 citations counted in INSPIRE as of 16 Jul 2018
%%%%%%%%%%%%%%%%%%%%%%%%%%%%%%%%%%%%%%%%%%%%%%%%%%%%%%%%%%%%%%%%%%%%%%%%%%%%%%%%%%%
%\cite{Herdeiro:2015kha}
\bibitem{Herdeiro:2015kha}
  C.~Herdeiro, J.~Kunz, E.~Radu and B.~Subagyo,
  ``Myers-Perry black holes with scalar hair and a mass gap: Unequal spins,''
  Phys.\ Lett.\ B {\bf 748} (2015) 30
  %doi:10.1016/j.physletb.2015.06.059
  [arXiv:1505.02407 [gr-qc]].
  %%CITATION = doi:10.1016/j.physletb.2015.06.059;%%
  %29 citations counted in INSPIRE as of 16 Jul 2018
%%%%%%%%%%%%%%%%%%%%%%%%%%%%%%%%%%%%%%%%%%%%%%%%%%%%%%%%%%%%%%%%%%%%%%%%%%%%%%%%%%%
%\cite{Herdeiro:2015tia}
\bibitem{Herdeiro:2015tia}
  C.~A.~R.~Herdeiro, E.~Radu and H.~Rœnarsson,
  ``Kerr black holes with self-interacting scalar hair: hairier but not heavier,''
  Phys.\ Rev.\ D {\bf 92} (2015) no.8,  084059
  %doi:10.1103/PhysRevD.92.084059
  [arXiv:1509.02923 [gr-qc]].
  %%CITATION = doi:10.1103/PhysRevD.92.084059;%%
  %66 citations counted in INSPIRE as of 16 Jul 2018
%%%%%%%%%%%%%%%%%%%%%%%%%%%%%%%%%%%%%%%%%%%%%%%%%%%%%%%%%%%%%%%%%%%%%%%%%%%%%%%%%%%
%\cite{Herdeiro:2016tmi}
\bibitem{Herdeiro:2016tmi}
  C.~Herdeiro, E.~Radu and H.~Runarsson,
  ``Kerr black holes with Proca hair,''
  Class.\ Quant.\ Grav.\  {\bf 33} (2016) no.15,  154001
  %doi:10.1088/0264-9381/33/15/154001
  [arXiv:1603.02687 [gr-qc]].
  %%CITATION = doi:10.1088/0264-9381/33/15/154001;%%
  %89 citations counted in INSPIRE as of 16 Jul 2018
  %%%%%%%%%%%%%%%%%%%%%%%%%%%%%%%%%%%%%%%%%%%%%%%%%%%%%%%%%%%%%%%%%%%%%%%%%%%%%%%%%%%
%\cite{Delgado:2016jxq}
\bibitem{Delgado:2016jxq}
  J.~F.~M.~Delgado, C.~A.~R.~Herdeiro, E.~Radu and H.~Runarsson,
  ``KerrÐNewman black holes with scalar hair,''
  Phys.\ Lett.\ B {\bf 761} (2016) 234
  %doi:10.1016/j.physletb.2016.08.032
  [arXiv:1608.00631 [gr-qc]].
  %%CITATION = doi:10.1016/j.physletb.2016.08.032;%%
  %18 citations counted in INSPIRE as of 16 Jul 2018
%%%%%%%%%%%%%%%%%%%%%%%%%%%%%%%%%%%%%%%%%%%%%%%%%%%%%%%%%%%%%%%%%%%%%%%%%%%%%%%%%%%
%\cite{Herdeiro:2018wvd}
\bibitem{Herdeiro:2018wvd}
  C.~A.~R.~Herdeiro and E.~Radu,
  %``Spinning boson stars and hairy black holes with non-minimal coupling,''
  arXiv:1803.08149 [gr-qc].
  %%CITATION = ARXIV:1803.08149;%%
%%%%%%%%%%%%%%%%%%%%%%%%%%%%%%%%%%%%%%%%%%%%%%%%%%%%%%%%%%%%%%%%%%%%%%%%%%%%%%%%%%%
%\cite{Herdeiro:2017oyt}
\bibitem{Herdeiro:2017oyt}
  C.~Herdeiro, J.~Kunz, E.~Radu and B.~Subagyo,
  ``Probing the universality of synchronised hair around rotating black holes with Q-clouds,''
  Phys.\ Lett.\ B {\bf 779} (2018) 151
  %doi:10.1016/j.physletb.2018.01.083
  [arXiv:1712.04286 [gr-qc]].
  %%CITATION = doi:10.1016/j.physletb.2018.01.083;%%
  %2 citations counted in INSPIRE as of 16 Jul 2018
%%%%%%%%%%%%%%%%%%%%%%%%%%%%%%%%%%%%%%%%%%%%%%%%%%%%%%%%%%%%%%%%%%%%%%%%%%%%%%%%%%%
%\cite{Herdeiro:2014pka}
\bibitem{Herdeiro:2014pka}
  C.~Herdeiro, E.~Radu and H.~Runarsson,
  ``Non-linear $Q$-clouds around Kerr black holes,''
  Phys.\ Lett.\ B {\bf 739} (2014) 302
%  doi:10.1016/j.physletb.2014.11.005
  [arXiv:1409.2877 [gr-qc]].
  %%CITATION = doi:10.1016/j.physletb.2014.11.005;%%
  %51 citations counted in INSPIRE as of 09 Jun 2018
 %%%%%%%%%%%%%%%%%%%%%%%%%%%%%%%%%%%%%%%%%%%%%%%%%%%%%%%%%%%
%\cite{Emparan:2001wn}
\bibitem{Emparan:2001wn}
  R.~Emparan and H.~S.~Reall,
  ``A rotating black ring solution in five-dimensions,''
  Phys.\ Rev.\ Lett.\  {\bf 88} (2002) 101101
  %doi:10.1103/PhysRevLett.88.101101
  [hep-th/0110260].
  %%CITATION = doi:10.1103/PhysRevLett.88.101101;%%
  %787 citations counted in INSPIRE as of 16 Jul 2018
%%%%%%%%%%%%%%%%%%%%%%%%%%%%%%%%%%%%%%%%%%%%%%%%%%%%%%%%%%%
     %\cite{Coleman:1985ki}
\bibitem{Coleman:1985ki}
  S.~R.~Coleman,
  ``Q Balls,''
  Nucl.\ Phys.\ B {\bf 262} (1985) 263
   Erratum: [Nucl.\ Phys.\ B {\bf 269} (1986) 744].
  %doi:10.1016/0550-3213(85)90286-X, 10.1016/0550-3213(86)90520-1
  %%CITATION = doi:10.1016/0550-3213(85)90286-X, 10.1016/0550-3213(86)90520-1;%%
  %713 citations counted in INSPIRE as of 19 Jul 2018
%%%%%%%%%%%%%%%%%%%%%%%%%%%%%%%%%%%%%%%%%%%%%%%%%%%%%%%%%%%
%\cite{Volkov:2002aj}
\bibitem{Volkov:2002aj}
  M.~S.~Volkov and E.~Wohnert,
  ``Spinning Q balls,''
  Phys.\ Rev.\ D {\bf 66} (2002) 085003
  %doi:10.1103/PhysRevD.66.085003
  [hep-th/0205157].
  %%CITATION = doi:10.1103/PhysRevD.66.085003;%%
  %69 citations counted in INSPIRE as of 08 Oct 2017
%%%%%%%%%%%%%%%%%%%%%%%%%%%%%%%%%%%%%%%%%%%%%%%%%%%%%%%%%%%%%%%%%%%%%%%%%%%%%%%%%%%
    %\cite{Kleihaus:2005me}
\bibitem{Kleihaus:2005me}
  B.~Kleihaus, J.~Kunz and M.~List,
  ``Rotating boson stars and Q-balls,''
  Phys.\ Rev.\ D {\bf 72} (2005) 064002
 % doi:10.1103/PhysRevD.72.064002
  [gr-qc/0505143].
  %%CITATION = doi:10.1103/PhysRevD.72.064002;%%
  %77 citations counted in INSPIRE as of 30 Sep 2017
  %%%%%%%%%%%%%%%%%%%%%%%%%%%%%%%%%%%%%%%%%%%%%%%%%%%%%%%%%%%
  %\cite{Hod:2014baa}
\bibitem{Hod:2014baa}
  S.~Hod,
  ``Kerr-Newman black holes with stationary charged scalar clouds,''
  Phys.\ Rev.\ D {\bf 90} (2014) no.2,  024051
 % doi:10.1103/PhysRevD.90.024051
  [arXiv:1406.1179 [gr-qc]].
  %%CITATION = doi:10.1103/PhysRevD.90.024051;%%
  %60 citations counted in INSPIRE as of 17 Jul 2018
   %%%%%%%%%%%%%%%%%%%%%%%%%%%%%%%%%%%%%%%%%%%%%%%%%%%%%%%%%%%
  %%%%%%%%%%%%%%%%%%%%%%%%%%%%%%%%%%%%%%%%%%%%%%%%%%%%%%%%%%%
%\cite{Benone:2014ssa}
\bibitem{Benone:2014ssa}
  C.~L.~Benone, L.~C.~B.~Crispino, C.~Herdeiro and E.~Radu,
  ``Kerr-Newman scalar clouds,''
  Phys.\ Rev.\ D {\bf 90} (2014) no.10,  104024
 % doi:10.1103/PhysRevD.90.104024
  [arXiv:1409.1593 [gr-qc]].
  %%CITATION = doi:10.1103/PhysRevD.90.104024;%%
  %75 citations counted in INSPIRE as of 09 Jun 2018
 %%%%%%%%%%%%%%%%%%%%%%%%%%%%%%%%%%%%%%%%%%%%%%%%%%%%%%%%%%%
  %%%%%%%%%%%%%%%%%%%%%%%%%%%%%%%%%%%%%%%%%%%%%%%%%%%%%%%%%%%
%\cite{Wilson-Gerow:2015esa}
\bibitem{Wilson-Gerow:2015esa}
  J.~Wilson-Gerow and A.~Ritz,
  ``Black hole energy extraction via a stationary scalar analog of the Blandford-Znajek mechanism,''
  Phys.\ Rev.\ D {\bf 93} (2016) no.4,  044043
  %doi:10.1103/PhysRevD.93.044043
  [arXiv:1509.06681 [hep-th]].
  %%CITATION = doi:10.1103/PhysRevD.93.044043;%%
  %8 citations counted in INSPIRE as of 17 Jul 2018
%%%%%%%%%%%%%%%%%%%%%%%%%%%%%%%%%%%%%%%%%%%%%%%%%%%%%%%%%%%
  %%%%%%%%%%%%%%%%%%%%%%%%%%%%%%%%%%%%%%%%%%%%%%%%%%%%%%%%%%%
%\cite{Bernard:2016wqo}
\bibitem{Bernard:2016wqo}
  C.~Bernard,
  ``Stationary charged scalar clouds around black holes in string theory,''
  Phys.\ Rev.\ D {\bf 94} (2016) no.8,  085007
 % doi:10.1103/PhysRevD.94.085007
  [arXiv:1608.05974 [gr-qc]].
  %%CITATION = doi:10.1103/PhysRevD.94.085007;%%
  %7 citations counted in INSPIRE as of 17 Jul 2018
%%%%%%%%%%%%%%%%%%%%%%%%%%%%%%%%%%%%%%%%%%%%%%%%%%%%%%%%%%%
  %%%%%%%%%%%%%%%%%%%%%%%%%%%%%%%%%%%%%%%%%%%%%%%%%%%%%%%%%%%
%\cite{Sakalli:2016xoa}
\bibitem{Sakalli:2016xoa}
  I.~Sakalli and G.~Tokgoz,
  ``Stationary scalar clouds around maximally rotating linear dilaton black holes,''
  Class.\ Quant.\ Grav.\  {\bf 34} (2017) no.12,  125007
  %doi:10.1088/1361-6382/aa6858
  [arXiv:1610.09329 [gr-qc]].
  %%CITATION = doi:10.1088/1361-6382/aa6858;%%
  %3 citations counted in INSPIRE as of 17 Jul 2018
%%%%%%%%%%%%%%%%%%%%%%%%%%%%%%%%%%%%%%%%%%%%%%%%%%%%%%%%%%%
  %%%%%%%%%%%%%%%%%%%%%%%%%%%%%%%%%%%%%%%%%%%%%%%%%%%%%%%%%%%
%\cite{Hod:2016lgi}
\bibitem{Hod:2016lgi}
  S.~Hod,
  ``Spinning Kerr black holes with stationary massive scalar clouds: The large-coupling regime,''
  JHEP {\bf 1701} (2017) 030
  %doi:10.1007/JHEP01(2017)030
  [arXiv:1612.00014 [hep-th]].
  %%CITATION = doi:10.1007/JHEP01(2017)030;%%
  %20 citations counted in INSPIRE as of 17 Jul 2018
%%%%%%%%%%%%%%%%%%%%%%%%%%%%%%%%%%%%%%%%%%%%%%%%%%%%%%%%%%%
  %%%%%%%%%%%%%%%%%%%%%%%%%%%%%%%%%%%%%%%%%%%%%%%%%%%%%%%%%%%
%\cite{Ferreira:2017cta}
\bibitem{Ferreira:2017cta}
  H.~R.~C.~Ferreira and C.~A.~R.~Herdeiro,
  ``Stationary scalar clouds around a BTZ black hole,''
  Phys.\ Lett.\ B {\bf 773} (2017) 129
  %doi:10.1016/j.physletb.2017.08.017
  [arXiv:1707.08133 [gr-qc]].
  %%CITATION = doi:10.1016/j.physletb.2017.08.017;%%
  %4 citations counted in INSPIRE as of 17 Jul 2018
%%%%%%%%%%%%%%%%%%%%%%%%%%%%%%%%%%%%%%%%%%%%%%%%%%%%%%%%%%%
  %%%%%%%%%%%%%%%%%%%%%%%%%%%%%%%%%%%%%%%%%%%%%%%%%%%%%%%%%%%
%\cite{Derrick:1964ww}
\bibitem{Derrick:1964ww}
  G.~H.~Derrick,
  ``Comments on nonlinear wave equations as models for elementary particles,''
  J.\ Math.\ Phys.\  {\bf 5} (1964) 1252.
  %doi:10.1063/1.1704233
  %%CITATION = doi:10.1063/1.1704233;%%
  %629 citations counted in INSPIRE as of 19 Jul 2018
%%%%%%%%%%%%%%%%%%%%%%%%%%%%%%%%%%%%%%%%%%%%%%%%%%%%%%%%%%%
\bibitem{Adam:2010fg}
  C.~Adam, J.~Sanchez-Guillen and A.~Wereszczynski,
``A Skyrme-type proposal for baryonic matter,''
  Phys.\ Lett.\ B {\bf 691} (2010) 105
%%%%%%%%%%%%%%%%%%%%%%%%%%%%%%%%%%%%%%%%%%%%%%%%%%%%%%%%%%%%%%%%%%%%%%%%%%%%%%
\bibitem{Adam:2010ds}
  C.~Adam, J.~Sanchez-Guillen and A.~Wereszczynski,
  ``A BPS Skyrme model and baryons at large $N_c$,''
  Phys.\ Rev.\ D {\bf 82} (2010) 085015
  %%%%%%%%%%%%%%%%%%%%%%%%%%%%%%%%%%%%%%%%%%%%%%%%%%%%%%%%%%%
%%%%%%%%%%%%%%%%%%%%%%%%%%%%%%%%%%%%%%%%%%%%%%%%%%%%%%%%%%%%%%%%%%%%%%%%%%%%%%
\bibitem{Perapechka:2017yyc}
  I.~Perapechka and Y.~Shnir,
``Crystal structures in generalized Skyrme model,''
  Phys.\ Rev.\ D {\bf 96} (2017) no.4,  045013
  %%%%%%%%%%%%%%%%%%%%%%%%%%%%%%%%%%%%%%%%%%%%%%%%%%%%%%%%%%%
    %\cite{Esteban:1986dm}
\bibitem{Esteban:1986dm}
  M.~J.~Esteban,
  ``A direct variational approach to Skyrme's model for meson fields,''
  Commun.\ Math.\ Phys.\  {\bf 105} (1986) 571.
 % doi:10.1007/BF01238934
  %%CITATION = doi:10.1007/BF01238934;%%
  %12 citations counted in INSPIRE as of 16 Jun 2018
%%%%%%%%%%%%%%%%%%%%%%%%%%%%%%%%%%%%%%%%%%%%%%%%%%%%%%%%%%%%%%%%%%%%%%%%%%%%%%%%%%%%%%%%%%%%%%%%%%%%%%%%%
%\cite{Manton:2004tk}
\bibitem{Manton:2004tk}
  N.~S.~Manton and P.~Sutcliffe,
  {\it 'Topological solitons',}
    Cambridge University Press, 2004.
  %%CITATION = INSPIRE-660150;%%
  %149 citations counted in INSPIRE as of 21 Sep 2017
%%%%%%%%%%%%%%%%%%%%%%%%%%%%%%%%%%%%%%%%%%%%%%%%%%%%%%%%%%%%%%%%%%%%%%%%%%%%%%%%%%%%
\bibitem{Krusch:2004uf}
  S.~Krusch and P.~Sutcliffe,
``Sphalerons in the Skyrme model,''
  J.\ Phys.\ A {\bf 37} (2004) 9037
%%%%%%%%%%%%%%%%%%%%%%%%%%%%%%%%%%%%%%%%%%%%%%%%%%%%%%%%%%%%%%%%%%%%%%%%%%%%%%%%%%%%
    %\cite{Shnir:2009ct}
\bibitem{Shnir:2009ct}
  Y.~Shnir and D.~H.~Tchrakian,
  ``Skyrmion-anti-Skyrmion chains,''
  J.\ Phys.\ A {\bf 43} (2010) 025401
  %doi:10.1088/1751-8113/43/2/025401
 % [arXiv:0906.5583 [hep-th]].
  %%CITATION = doi:10.1088/1751-8113/43/2/025401;%%
  %6 citations counted in INSPIRE as of 16 Jun 2018
    %%%%%%%%%%%%%%%%%%%%%%%%%%%%%%%%%%%%%%%%%%%%%%%%%%%%%%%%%%%%%%%%%%%%%%%%%%%%%%%%%%%%
    %\cite{Smolic:2015txa}
\bibitem{Smolic:2015txa}
  I.~Smolic,
  ``Symmetry inheritance of scalar fields,''
  Class.\ Quant.\ Grav.\  {\bf 32} (2015) no.14,  145010
  %doi:10.1088/0264-9381/32/14/145010
  [arXiv:1501.04967 [gr-qc]].
  %%CITATION = doi:10.1088/0264-9381/32/14/145010;%%
  %19 citations counted in INSPIRE as of 27 Jul 2018
 %%%%%%%%%%%%%%%%%%%%%%%%%%%%%%%%%%%%%%%%%%%%%%%%%%%%%%%%%%%%%%%%%%%%%%%%%%%%%%%%%%%%
    %\cite{BoutalebJoutei:1979va}
\bibitem{BoutalebJoutei:1979va}
  H.~Boutaleb-Joutei, A.~Chakrabarti and A.~Comtet,
  ``Gauge field configurations in curved space-times,''
  Phys.\ Rev.\ D {\bf 20} (1979) 1884.
 % doi:10.1103/PhysRevD.20.1884
  %%CITATION = doi:10.1103/PhysRevD.20.1884;%%
  %31 citations counted in INSPIRE as of 16 Jun 2018
    %%%%%%%%%%%%%%%%%%%%%%%%%%%%%%%%%%%%%%%%%%%%%%%%%%%%%%%%%%%%%%%%%%%%%%%%%%%%%%%%%%%%
 \bibitem{Gudnason:2018lhn}
  S.~B.~Gudnason and M.~Nitta,
``Higher-order Skyrme hair of black holes,''
  JHEP {\bf 1805} (2018) 071
%%%%%%%%%%%%%%%%%%%%%%%%%%%%%%%%%%%%%%%%%%%%%%%%%%%%%%%%%%%
\bibitem{pardiso}
O.~Schenk and K-~G\"artner
``Solving unsymmetric sparse systems of linear equations with PARDISO,``
Future Generation Computer Systems, 20(3) (2004) 475
 %%%%%%%%%%%%%%%%%%%%%%%%%%%%%%%%%%%%%%%%%%%%%%%%%%%%%%%%%%%
\bibitem{schoen}
 W. Sch\"onauer and R. Wei\ss ,
``Efficient vectorizable PDE solvers,''
 J. Comput. Appl. Math. 27, 279 (1989) 279;
 \\
 M. Schauder, R. Wei\ss\ and W. Sch\"onauer,
``The CADSOL Program Package,``
 Universit\"at Karlsruhe, Interner Bericht Nr. 46/92 (1992).
 %%%%%%%%%%%%%%%%%%%%%%%%%%%%%%%%%%%%%%%%%%%%%%%%%%%%%%%%%%%
\bibitem{Rajaraman}
    R. Rajaraman,
{\it Solitons and Instantons: An Introduction to Solitons and Instantons in Quantum Field Theory},
North-Holland Publishing Company, 1982
%%%%%%%%%%%%%%%%%%
%%%%%%%%%%%%%%%%%%%%%%%%%%%%%%%%%%%%%%%%%%%%%%%%%%%%%%%%%%%
    %\cite{Radu:2008pp}
\bibitem{Radu:2008pp}
  E.~Radu and M.~S.~Volkov,
  ``Existence of stationary, non-radiating ring solitons in field theory: knots and vortons,''
  Phys.\ Rept.\  {\bf 468} (2008) 101
 % doi:10.1016/j.physrep.2008.07.002
  [arXiv:0804.1357 [hep-th]].
  %%CITATION = doi:10.1016/j.physrep.2008.07.002;%%
  %78 citations counted in INSPIRE as of 27 May 2018
     %%%%%%%%%%%%%%%%%%%%%%%%%%%%%%%%%%%%%%%%%%%%%%%%%%%%%%%%%%%
     %%%%%%%%%%%%%%%%%%%%%%%%%%%%%%%%%%%%%%%%%%%%%%%%%%%%%%%%%%%
%\cite{Delgado:2016zxv}
\bibitem{Delgado:2016zxv}
  J.~F.~M.~Delgado, C.~A.~R.~Herdeiro and E.~Radu,
  ``Violations of the Kerr and Reissner-Nordstr\"om bounds: Horizon versus asymptotic quantities,''
  Phys.\ Rev.\ D {\bf 94} (2016) no.2,  024006
%  doi:10.1103/PhysRevD.94.024006
  [arXiv:1606.07900 [gr-qc]].
  %%CITATION = doi:10.1103/PhysRevD.94.024006;%%
  %5 citations counted in INSPIRE as of 26 Jun 2018
  %%%%%%%%%%%%%%%%%%%%%%%%%%%%%%%%%%%%%%%%%%%%%%%%%%%%%%%%%%%
    %\cite{Volkov:1998cc}
\bibitem{Volkov:1998cc}
  M.~S.~Volkov and D.~V.~Gal'tsov,
  ``Gravitating non-Abelian solitons and black holes with Yang-Mills fields,''
  Phys.\ Rept.\  {\bf 319} (1999) 1
 % doi:10.1016/S0370-1573(99)00010-1
  [hep-th/9810070].
  %%CITATION = doi:10.1016/S0370-1573(99)00010-1;%%
  %294 citations counted in INSPIRE as of 21 Sep 2017
  %%%%%%%%%%%%%%%%%%%%%%%%%%%%%%%%%%%%%%%%%%%%%%%%%%%%%%%%%%%
%\cite{Brandt:1996si}
\bibitem{Brandt:1996si}
  S.~R.~Brandt and E.~Seidel,
  %``The Evolution of distorted rotating black holes. 3: Initial data,''
  Phys.\ Rev.\ D {\bf 54} (1996) 1403
%  doi:10.1103/PhysRevD.54.1403
  [gr-qc/9601010].
  %%CITATION = doi:10.1103/PhysRevD.54.1403;%%
  %55 citations counted in INSPIRE as of 06 Oct 2018		
%%%%%%%%%%%%%%%%%%%%%%%%%%%%%%%%%%%%%%%%%%%%%%%%%%%%%%%%%%%%%%%%%%%%%%%%%%%%%%%%%%%%%%%%%%%%%%%%%%%%%%%%%%
%\cite{Yoshida:1997qf}
\bibitem{Yoshida:1997qf}
  S.~Yoshida and Y.~Eriguchi,
  ``Rotating boson stars in general relativity,''
  Phys.\ Rev.\ D {\bf 56} (1997) 762.
  %doi:10.1103/PhysRevD.56.762
  %%CITATION = doi:10.1103/PhysRevD.56.762;%%
  %68 citations counted in INSPIRE as of 07 Oct 2017
%%%%%%%%%%%%%%%%%%%%%%%%%%%%%%%%%%%%%%%%%%%%%%%%%%%%%%%%%%%%%%%%%%%
%\cite{Brihaye:2017wqa}
\bibitem{Brihaye:2017wqa}
  Y.~Brihaye, C.~Herdeiro, E.~Radu and D.~H.~Tchrakian,
  ``Skyrmions, Skyrme stars and black holes with Skyrme hair in five spacetime dimension,''
  JHEP {\bf 1711} (2017) 037
 % doi:10.1007/JHEP11(2017)037
  [arXiv:1710.03833 [gr-qc]].
  %%CITATION = doi:10.1007/JHEP11(2017)037;%%
  %2 citations counted in INSPIRE as of 16 Jul 2018
%%%%%%%%%%%%%%%%%%%%%%%%%%%%%%%%%%%%%%%%%%%%%%%%%%%%%%%%%%%%%%%%%%%%%%%%%%%%%%%%%%%








 

\end{thebibliography}
\end{document}